\documentclass{article}
\usepackage{amssymb}

\usepackage{amsmath}

\newtheorem{theorem}{Theorem}

\newtheorem{corollary}[theorem]{Corollary}

\newtheorem{definition}[theorem]{Definition}

\newtheorem{lemma}[theorem]{Lemma}

\newtheorem{proposition}[theorem]{Proposition}
\newtheorem{remark}[theorem]{Remark}

\newenvironment{proof}[1][Proof]{\noindent\textbf{#1.} }{\ \rule{0.5em}{0.5em}}
\input{tcilatex}

\begin{document}

\title{Witten Laplacian Methods for The Decay of Correlations}
\author{Assane Lo \\
The University of Arizona}
\maketitle

\begin{abstract}
The aim of this paper is to apply direct methods to the study of integrals
that appear naturally in Statistical Mechanics and Euclidean Field Theory.
We provide weighted estimates leading to the exponential decay of the
two-point correlation functions for certain classical convex unbounded
models. The methods involve the study of the solutions of the Witten
Laplacian equations associated with the Hamiltonian of the system.
\end{abstract}

\section{Introduction}

In these notes, we study partial differential equation techniques for
problems coming from equilibrium Statistical Mechanics and Euclidean Field
theory. In the context of classical equilibrium Statistical Mechanics, one
is interested in a natural mathematical description of an equilibrium state
of a physical system which consists of a very large number of interacting
components. Consider, for example a piece of ferromagnetic metal (like iron,
cobalt, or nickel) in thermal equilibrium. The piece consists of a very
large number of atoms which are located at the sites of a crystal lattice $%
\Lambda .$ Each atom shows a magnetic moment which can be visualized as a
vector in $\mathbb{R}^{3}.$ This magnetic moment is called the \textit{spin }%
of the atom and represents the orientation of the atom in the lattice. The
set $S$ of all possible orientations of the spins, is called the \textit{%
state space} of the system. Each element $i$ of $\Lambda $ is called a
(lattice) \textit{site.} A particular configuration of the total system will
be described by an element $x=\left( x_{i}\right) _{i\in \Lambda }$ of the
product space $\Omega =S^{\Lambda }.$ This set $\Omega $ is called the
configuration space.\newline
The physical system considered above is characterized by a sharp contrast:
the microscopic structure is enormously complex, and any measurement of
microscopic quantities is subject to Statistical fluctuations. The
macroscopic behavior, however, can be described by means of a few parameters
such as magnetization and temperature, and macroscopic measurement leads to
apparently deterministic results. This contrast between the microscopic and
the macroscopic level is the starting point of Classical Statistical
Mechanics as developed by Maxwell, Boltzman, and Gibbs. Their basic idea may
be summarized as follows: The microscopic complexity may be overcome by a
statistical approach, and the macroscopic determinism then may be regarded
as a consequence of a suitable law of large numbers. According to this
philosophy, it is not adequate to describe the state of the system by a
particular element $x$ of the configuration space $\Omega .$ The system's
state should rather be described by a family of $S-$valued random variables
or (if we pass to the joint distribution of these random variables), by a
probability measure $\mu $ on $\Omega $ consistent with the available
partial knowledge of the system. In particular, $\mu $ should take account
of the a priory assumption that the system is in thermal equilibrium.\newline
Which kind of probability measure on $\Omega $ is suitable to describe a
physical system in equilibrium? The term equilibrium clearly refers to the
notion of forces and energies that act on the system. Thus one needs to
define a Hamiltonian $\Phi $ which assigns to each configuration $x$ a
potential energy $\Phi (x).$ In the physical system above, the essential
contribution to the potential energy comes from the interaction of the
microscopic components of the system and a possible external force. As soon
as a Hamiltonian $\Phi $ have been specified, the answer to the question is
generally believed to be the probability measure 
\begin{equation*}
d\mu (x)=Z^{-1}e^{-\beta \Phi (x)}d\lambda (x).
\end{equation*}%
Here $d\lambda $ refers to a suitable a priory measure (for example the
counting measure if $\Omega $ is finite), $\beta $ is a positive number
which is proportional to the inverse of the absolute temperature and $Z>0$
is a normalization constant. The above measure $\mu $ is called \textit{the
Boltzmann-Gibbs distribution}.\newline
As we have mentioned above the number of atoms in a ferromagnet is extremely
large. Consequently, the set $\Lambda $ in our mathematical model should be
very large. According to a standard rule of a mathematical thinking, the
intrinsic properties of large objects can be made manifest by performing
suitable limiting procedures. It is therefore a common practice in
Statistical Physics to pass to the infinite volume limit $\left| \Lambda
\right| \rightarrow \infty .$ (This limit is also referred to as the
thermodynamic limit). The Boltzmann-Gibbs distribution does not admit a
direct extension to infinite systems. However, when dealing with infinite
systems, we can still look at finite subsystems provided the rest is held
fixed. Indeed, starting with an interacting potential $\phi $ we can define
for each finite subsystem $\Lambda $ a Hamiltonian $\Phi _{\Lambda }^{\phi }$
which includes the interactions of $\Lambda $ with its fixed environment.%
\newline
The methods for investigating phase transition for certain physical systems
took an interesting direction when powerful and sophisticated PDE techniques
are introduced in the mathematical technology. The methods are generally
based on the analysis of suitable differential operators%
\begin{equation*}
\mathbf{W}_{\Phi }^{\left( 0\right) }=\left( \mathbf{-\Delta +}\frac{\left| 
\mathbf{\nabla }\Phi \right| ^{2}}{4}-\frac{\mathbf{\Delta }\Phi }{2}\right)
\end{equation*}%
and%
\begin{equation*}
\mathbf{W}_{\Phi }^{\left( 1\right) }=\mathbf{-\Delta +}\frac{\left| \mathbf{%
\nabla }\Phi \right| ^{2}}{4}-\frac{\mathbf{\Delta }\Phi }{2}+\mathbf{Hess}%
\Phi .
\end{equation*}%
These are in some sense, deformations of the standard Laplace Beltrami
operator. These operators, commonly called Witten Laplacians, were first
introduced by Edward Witten [18] in 1982 in the context of Morse theory for
the study of topological invariants of compact Riemannian manifolds. In
1994, Bernard Helffer and J\"{o}hannes Sj\"{o}strand [8] introduced two
elliptic differential operators.%
\begin{equation*}
A_{\Phi }^{(0)}:=-\mathbf{\Delta }+\mathbf{\nabla }\Phi \cdot \mathbf{\nabla 
}
\end{equation*}%
and 
\begin{equation*}
A_{\Phi }^{(1)}:=-\mathbf{\Delta }+\mathbf{\nabla }\Phi \cdot \mathbf{\nabla
+Hess}\Phi .
\end{equation*}%
These later operators, provide direct methods for the study of integrals and
operators in high dimensions of the type that appear in statistical
mechanics and euclidean field theory. In 1996, J. Sj\"{o}strand [13]
observed that these so called Helffer-Sj\"{o}strand operators are in fact
equivalent to Witten's Laplacians. Since then, there have been significant
advances in the use of these Laplacians to study the thermodynamic behavior
of quantities related to the Gibbs measure $Z^{-1}e^{-\Phi }dx.$ As a simple
illustration, if one is interested in the study of the mean value $%
\left\langle g\right\rangle _{\Lambda }$ where 
\begin{equation*}
\left\langle g\right\rangle _{\Lambda }=\int gd\mu _{\Lambda }
\end{equation*}%
and 
\begin{equation*}
d\mu _{\Lambda }=\frac{e^{-\Phi _{\Lambda }}dx}{\int e^{\Phi _{\Lambda }}dx}
\end{equation*}%
for a suitable smooth function $g,$ one can first solve the equation%
\begin{equation*}
\mathbf{\nabla }g=\left( -\mathbf{\Delta }+\mathbf{\nabla }\Phi \cdot 
\mathbf{\nabla }\right) \mathbf{v+Hess}\Phi \mathbf{v,}
\end{equation*}%
for a $C^{\infty }-$solution $\mathbf{v}$ where the operator 
\begin{equation*}
-\mathbf{\Delta }+\mathbf{\nabla }\Phi \cdot \mathbf{\nabla }
\end{equation*}%
acts diagonally on each component of $\mathbf{v.}$ Under suitable
assumptions on the Hamiltonian $\Phi ,$ one can see that $\mathbf{v}$ is
also a solution of the system%
\begin{equation*}
g=<g>_{\Lambda }+\mathbf{v}\cdot \mathbf{\nabla }\Phi -div\mathbf{v.}
\end{equation*}%
If it turns out that $g(0)=0$ and $0$ is a critical point of $\Phi ,$ then 
\begin{equation*}
<g>_{\Lambda }=div\mathbf{v}(0).
\end{equation*}%
Thus, the study of the thermodynamic properties of the mean value is then
reduced to estimating the derivatives of the solution $\mathbf{v.}$

One of the most striking results is an exact formula for the covariance of
two functions in terms of the Witten Laplacian on one forms, leading to
sophisticated methods for estimating the correlation functions. This formula
is in some sense a stronger and more flexible version of the Brascamp-Lieb
inequality [1]. The formula may be written as follow:%
\begin{equation}
\mathbf{cov}(g,h)=\int \left( A_{\Phi }^{(1)^{-1}}\mathbf{\nabla }g\cdot 
\mathbf{\nabla }h\right) e^{-\Phi (x)}dx.
\end{equation}%
To understand the idea behind this formula, let us denote by $\left\langle
f\right\rangle $ the mean value of $f$ with respect to the measure 
\begin{equation*}
e^{-\Phi (x)}dx,
\end{equation*}%
the covariance of two functions $f$ and $g$ is defined by 
\begin{equation}
\mathbf{cov}(g,h)=\left\langle (g-\left\langle g\right\rangle
)(h-\left\langle h\right\rangle )\right\rangle .
\end{equation}%
If one wants to have an expression of the covariance in the form 
\begin{equation}
\mathbf{cov}(g,h)=\left\langle \mathbf{\nabla }h\cdot \mathbf{w}%
\right\rangle _{L^{2}(\mathbb{R}^{n},\mathbb{R}^{n};e^{-\Phi }dx)},
\end{equation}%
for a suitable vector field $\mathbf{w,}$ we get, after observing that $%
\mathbf{\nabla }h=\mathbf{\nabla }(h-\left\langle h\right\rangle ),$ and
integrating by parts:%
\begin{equation}
\mathbf{cov}(g,h)=\int (h-\left\langle h\right\rangle )(\mathbf{\nabla }\Phi
-\mathbf{\nabla })\cdot \mathbf{w}e^{-\Phi (x)}dx.
\end{equation}%
This leads to the question of solving the equation%
\begin{equation}
g-\left\langle g\right\rangle =(\mathbf{\nabla }\Phi -\mathbf{\nabla })\cdot 
\mathbf{w.}
\end{equation}%
Now trying to solve this above equation with $\mathbf{w}=\mathbf{\nabla }u,$
we obtain the equation%
\begin{equation}
\left. 
\begin{array}{c}
g-\left\langle g\right\rangle =A_{\Phi }^{(0)}u \\ 
\left\langle u\right\rangle =0.%
\end{array}%
\right\}
\end{equation}%
Assuming for now the existence of a smooth solution, we get by
differentiation of this above equation 
\begin{equation}
\mathbf{\nabla }g=A_{\Phi }^{(1)}\mathbf{\nabla }u
\end{equation}%
and the formula is now easy to see.

New methods that are purely based on spectral analysis have been recently
developed by Helffer-Bodineau [2], Sjostrand-Bach-Jecko [26]. In these
papers, the authors studied a certain class of unbounded spin models by
means of the spectra of the Witten Laplacian. In [26], the asymptotics of
the two point correlation function to leading order in $\beta ^{-1}$ was
obtained under under weaker assumptions on the Hamiltonian. In 2003, V. Bach
and J. S. Moller [27] proposed a refined version of the results in [26] by
introducing a new twisted Witten Laplacian to relax the convexity
assumptions.

We attempt in this paper, to study weighted estimates that lead to the
exponential decay of the two-point correlation functions for certain convex
unbounded systems. We removed limitations of earlier work of Helffer and Sj%
\"{o}strand [8].They only treated the one dimensional case ($d=1$) under the
artificial restrictions%
\begin{equation*}
\left\| \mathbf{Hess}\Phi (x)\right\| _{\mathcal{L}(l_{\rho }^{\infty
})}\leq C
\end{equation*}%
and 
\begin{equation*}
\left\| \mathbf{Hess}\Phi (x)-\mathbf{I}\right\| _{\mathcal{L}(l_{\rho
}^{\infty })}\leq \delta <1,
\end{equation*}%
\ \ for all weight function $\rho $ on $\mathbb{Z}/m\mathbb{Z}$ satisfying%
\begin{equation*}
e^{-\kappa }\leq \frac{\rho \left( i+1\right) }{\rho (i)}\leq e^{\kappa },\ 
\text{for some }\kappa >0.
\end{equation*}%
These conditions are too restrictive for many important applications, while
my conditions are considerably more flexible. In particular, the conditions
in my work are suitable for treating the d-dimensional nearest neighbor Kac
model, where the potential is given by 
\begin{equation*}
\Phi (x)=\frac{x^{2}}{2}-2\sum_{i,j\in \Lambda ,i\sim j}\ln \cosh \left[ 
\sqrt{\frac{\nu }{2}}\left( x_{i}+x_{j}\right) \right] ,\;\;\;\;\;x=\left(
x_{i}\right) _{i\in \Lambda },
\end{equation*}%
for $\nu >0$ smaller than some value $\nu _{o}$ to be determined.

In section 2, we give a motivational background on the origin of Witten's
Laplacians.

In section 3, we give an outline of the operators and equations involved in
the Witten Laplacian method.

In section 4, we discuss preliminary results on Hilbert space methods for
elliptic PDE's.

In section 5, we provide a rigorous discussion based on Hilbert space
methods for the solvability of the corresponding Witten Laplacian equations.

In section 6, we illustrate the family of Hamiltonians discussed in section
3 and 5 through an example of the type introduced by Marc Kac [20]. Section
7 is devoted to the study of the\ exponential decay of the two-point
correlation functions for models of Kac type in the convex case. We shall
establish weighted estimates leading to the exponential decay of the
two-point correlation functions.

In section 8, we shall apply our methods to the d-dimensional nearest
neighbor Kac model.

\section{The Witten's Laplacians}

In 1982, Edward Witten published an article [18] on Supersymmetry and Morse
theory relating invariants of a Riemannian manifold $\mathbf{M}$ with some
indices of a Morse function $\Phi \in C^{\infty }(\mathbf{M}).$ For this, he
introduced the Witten derivative $\mathbf{d}_{\Phi }$ and the Witten
coderivative $\mathbf{d}_{\Phi }^{\ast }$ by simply setting$\;$%
\begin{equation}
\mathbf{d}_{\Phi }\mathbf{=e}^{-\frac{\Phi }{2}}\mathbf{de}^{\frac{\Phi }{2}}%
\mathbf{\;\ \ \ }\text{and}\ \ \ \ \ \mathbf{d}_{\Phi }^{\ast }\mathbf{=e}^{%
\frac{\Phi }{2}}\mathbf{d}^{\ast }\mathbf{e}^{-\frac{\Phi }{2}}\;\;,
\end{equation}%
where $\mathbf{d}$ and $\mathbf{d}^{\ast }$ are the exterior derivative and
exterior coderivative respectively. The Witten Laplacian is then defined to
be the associated second order operator 
\begin{eqnarray}
\mathbf{W}_{\Phi } &=&\left( \mathbf{d}_{\Phi }+\mathbf{d}_{\Phi }^{\ast
}\right) ^{2} \\
&=&\mathbf{d}_{\Phi }\mathbf{d}_{\Phi }^{\ast }+\mathbf{d}_{\Phi }^{\ast }%
\mathbf{d}_{\Phi }
\end{eqnarray}%
acting on the exterior algebra bundle of the cotangent bundle of $M$\textbf{%
\ }as the standard Laplacian does$.$

Choosing a local orthonormal frame field $\mathbf{e}_{1},...,\mathbf{e}_{d}$
and denoting by $\mathbf{e}^{1},...,\mathbf{e}^{d}$ its dual coframe field, $%
\mathbf{d}$ and $\mathbf{d}^{\ast }$ could be easily represented in terms of
the Riemannian connection $\mathbf{\nabla }$\textbf{\ }as 
\begin{equation}
\mathbf{d}=\mathbf{e}^{i}\mathbf{\wedge \nabla }_{e_{i}}\text{ \ \ \ \ and \
\ \ \ }\mathbf{d}^{\ast }=-\mathbf{\hat{\imath}}_{\mathbf{e}_{j}}\mathbf{%
\nabla }_{e_{j}.}
\end{equation}%
where $\mathbf{\hat{\imath}}_{\mathbf{e}_{j}}$ denote the interior product
with respect to $e_{j}$ (see [58] for more details). Here and in the rest of
this section, we use the Einstein summation convention namely, an index
occurring twice in a product is to be summed from $1$ up to the space
dimension. We consequently have 
\begin{equation}
\mathbf{d}_{\Phi }=\mathbf{e}^{i}\mathbf{\wedge \nabla }_{e_{i}}+\mathbf{e}%
^{i}\frac{\Phi _{;\;i}}{2}\;\ \ \ \text{and \ \ \ }\mathbf{d}_{\Phi }^{\ast
}=-\mathbf{i(e}_{j}\mathbf{)\nabla }_{e_{j}}+\mathbf{i(e}_{j}\mathbf{)}\frac{%
\Phi _{;\;i}}{2}
\end{equation}%
where $\Phi _{;\;i_{1}i_{2}...}$ denote the components of multiple covariant
differentiation relative to the local frame field $\mathbf{e}_{1},...,%
\mathbf{e}_{d}.$%
\begin{equation}
\Phi _{;\;ij}=\mathbf{\nabla }_{\mathbf{e}_{j}}\mathbf{\nabla _{\mathbf{e}%
_{i}}}\Phi \mathbf{-\nabla }_{\nabla _{\mathbf{e}_{j}}\mathbf{e}_{i}.}\Phi
\end{equation}%
Since $\mathbf{e}^{i}\mathbf{\wedge \nabla }_{e_{i}}$ and $\mathbf{i(e}_{j}%
\mathbf{)\nabla }_{e_{j}}$ do not depend on the choice of the local
orthonormal frame and coframe field we may assume that $\mathbf{e}_{1},...,%
\mathbf{e}_{d}$ comes from a normal coordinate centered at an arbitrary
point, and consequently have 
\begin{equation}
\mathbf{\nabla _{\mathbf{e}_{j}}e}^{i}\mathbf{\wedge =\nabla }_{\mathbf{e}%
_{i}}\mathbf{i(e}_{j}\mathbf{)=0.}
\end{equation}%
Now using (10), (11), (14) and the fact that 
\begin{equation}
\mathbf{e}^{i}\wedge \mathbf{i(e}_{j}\mathbf{)+i(e}_{j}\mathbf{)e}^{i}\wedge
=\delta _{ij},
\end{equation}%
we have%
\begin{equation}
\mathbf{W}_{\Phi }^{\left( p\right) }\mathbf{=\Delta +}\frac{\Phi
_{;\;i}\Phi _{;\;i}}{4}\mathbf{+}\frac{\Phi _{;\;ij}}{2}\mathbf{(e}^{i}%
\mathbf{\wedge i(e}_{j}\mathbf{)-i(e}_{j}\mathbf{)e}^{i}\mathbf{\wedge ).}
\end{equation}%
In the case of $\mathbb{R}^{n}$ where covariant differentiation becomes
standard differentiation, the Witten Laplacian on 0-forms acting on a smooth
function $f$ gives%
\begin{eqnarray}
\mathbf{W}_{\Phi }^{(0)}f &\mathbf{=}&\mathbf{-\Delta }f\mathbf{+}\frac{\Phi
_{x_{i}}\Phi _{x_{i}}}{4}f\mathbf{-}\frac{\Phi _{x_{i}x_{i}}}{2}f \\
&=&\left( \mathbf{-\Delta +}\frac{\left| \mathbf{\nabla }\Phi \right| ^{2}}{4%
}-\frac{\mathbf{\Delta }\Phi }{2}\right) f.
\end{eqnarray}%
\newline
The Witten Laplacian on one-forms acting on a one form 
\begin{equation*}
u=u^{k}(x)dx^{k}
\end{equation*}%
gives%
\begin{equation}
\mathbf{W}_{\phi }^{\left( 1\right) }u\mathbf{=\Delta }u\mathbf{+}\frac{%
\mathbf{\phi }_{x_{i}}\mathbf{\phi }_{x_{i}}}{4}u\mathbf{-}\frac{\mathbf{%
\phi }_{x_{i}x_{i}}}{2}u+2\frac{\mathbf{\phi }_{x_{k}x_{i}}}{2}dx^{i}\mathbf{%
\wedge i}_{\tfrac{\partial }{\partial x_{k}}}u.
\end{equation}%
Identifying one-forms with vector fields in $\mathbb{R}^{n}$ (1.12) becomes 
\begin{equation}
\mathbf{W}_{\Phi }^{\left( 1\right) }\mathbf{u=}\left( \mathbf{-\Delta +}%
\frac{\left| \mathbf{\nabla }\Phi \right| ^{2}}{4}-\frac{\mathbf{\Delta }%
\Phi }{2}\right) \otimes \mathbf{u}+\mathbf{Hess}\Phi \mathbf{u}.
\end{equation}%
The tensor notation simply means that the operator $\mathbf{-\Delta +}\dfrac{%
\left| \mathbf{\nabla }\Phi \right| ^{2}}{4}-\dfrac{\mathbf{\Delta }\Phi }{2}
$ acts diagonally on each component of the vector field $\mathbf{u.}$ Let us
also point out that the identification between forms and vector fields is a
common practice in Riemaniann geometry and is done via the metric tensor.

As first observed in [8] by Bernard Helffer and Johannes Sj\"{o}strand,
these Laplacians provide new methods for solving problems coming from
Statistical Mechanics. The methods are generally based on the analysis of
the differential operators 
\begin{equation}
A_{\Phi }^{(0)}:=-\mathbf{\Delta }+\mathbf{\nabla }\Phi \cdot \mathbf{\nabla 
}
\end{equation}%
and%
\begin{equation}
A_{\Phi }^{(1)}:=A_{\Phi }^{(0)}\otimes Id\mathbf{+Hess}\Phi .
\end{equation}%
These two elliptic differential operators for which a Fredholm theory can be
developed are equivalent, as observed in [13], to Witten's Laplacians $%
W_{\Phi }^{(0)}$ and $W_{\Phi }^{(1)}$ respectively where%
\begin{equation}
\mathbf{W}_{\Phi }^{\left( 0\right) }\mathbf{=-\Delta +}\frac{\left| \mathbf{%
\nabla }\Phi \right| ^{2}}{4}-\frac{\mathbf{\Delta }\Phi }{2}
\end{equation}%
and%
\begin{equation}
\mathbf{W}_{\Phi }^{\left( 1\right) }=\left( \mathbf{-\Delta +}\frac{\left| 
\mathbf{\nabla }\Phi \right| ^{2}}{4}-\frac{\mathbf{\Delta }\Phi }{2}\right)
\otimes \mathbf{I}+\mathbf{Hess}\Phi .
\end{equation}%
Indeed, it only suffices to observe that 
\begin{equation}
W_{\Phi }^{(.)}=e^{-\Phi /2}\circ A_{\Phi }^{(.)}\circ e^{\Phi /2}
\end{equation}%
and the map 
\begin{eqnarray*}
U_{\Phi } &:&L^{2}(\mathbb{R}^{\Lambda })\rightarrow L^{2}(\mathbb{R}%
^{\Lambda },e^{-\Phi }dx) \\
u &\longmapsto &e^{\frac{\Phi }{2}}u
\end{eqnarray*}%
is unitary.

\section{The Basic Equation}

For any\ finite domain $\Lambda $ of $\mathbb{Z}^{d},$ we shall consider a
Hamiltonian $\Phi _{\Lambda }$ of the phase space $\mathbb{R}^{\Lambda },$
satisfying conditions that will guaranty the solvability of the
corresponding Witten Laplacian equations. We shall also consider a slowly
growing source term $g,$ to ensure that the solutions have suitable
asymptotic behavior.

We shall first establish the solvability of the equation%
\begin{equation}
\left\{ 
\begin{tabular}{l}
$A_{\Phi }^{(0)}v=g-\left\langle g\right\rangle _{_{L^{2}(\mu )}}$ \\ 
$\left\langle v\right\rangle _{L^{2}(\mu )}=0$%
\end{tabular}%
\right.
\end{equation}%
by means of Hilbert space methods. The method consists of determining an
appropriate function space and an operator which is a natural realization of
the problem. In this particular problem, the function spaces to be
considered are the Sobolev spaces $B_{\Phi }^{k}(\mathbb{R}^{\Lambda })$
defined by 
\begin{equation*}
B_{\Phi }^{k}(\mathbb{R}^{\Lambda })=\left\{ u\in L^{2}(\mathbb{R}^{\Lambda
}):Z_{\Phi }^{\ell }\partial ^{\alpha }u\in L^{2}(\mathbb{R}^{\Lambda
})\;\forall \;\ell +\left| \alpha \right| \leq k\right\} .
\end{equation*}%
where 
\begin{equation}
Z_{\Phi }=\frac{\left| \mathbf{\nabla }\Phi \right| }{2}
\end{equation}%
These are subspaces of the well known Sobolev spaces $W^{k,2}(\mathbb{R}%
^{\Lambda }),$ $k$ $\in \mathbb{N}$.

The vital tool in the Hilbert space approach to elliptic boundary value
problems is the celebrated Lax-Milgram theorem. The essence of the method is
the interpretation of the problem in a variational sense involving a
bilinear form defined in a natural way by the problem and acting on the
appropriately chosen function spaces.

In general, the Hilbert space method for elliptic differential equations
uses the Compact embedding theorem for Sobolev spaces. This is a fundamental
step in the method in order to be able to apply the Fredholm alternative.
Since, in the context of our problem we are dealing with unbounded domains,
the classical results regarding the compactness of the embedding 
\begin{equation}
W^{k,p}(\Omega ,dx)\hookrightarrow L^{p}(\Omega ,dx)
\end{equation}%
for suitable $\Omega $ are no longer valid. However, In the case where the $%
L^{p}$ spaces are taken with respect to the weighted measure $e^{-\Phi }dx,$
with a suitable $\Phi ,$ we have the following result due to J-M$.$ Kneib
and F. Mignot [11]lem.5.

\begin{lemma}
\textit{If }$\Phi $\textit{\ satisfies the condition }%
\begin{equation*}
\exists \theta \in (0,1):\;\lim\limits_{\left| x\right| \rightarrow \infty
}\theta \left| \mathbf{\nabla }\Phi (x)\right| ^{2}-\mathbf{\Delta }\Phi
=\infty
\end{equation*}%
\textit{then }%
\begin{equation*}
H^{1}(\mu )\hookrightarrow L^{2}(\mathbb{R}^{\Lambda },d\mu )
\end{equation*}%
\textit{is compact.}
\end{lemma}

Here and in the sequel, $d\mu $ \ will denote the Gibbs measure%
\begin{equation*}
d\mu =Z^{-1}e^{-\Phi }dx,
\end{equation*}%
\begin{equation*}
Z=\int_{\mathbb{R}^{\Lambda }}e^{-\Phi }dx.
\end{equation*}%
and $H^{k}(\mu )$ denotes the weighted Sobolev space 
\begin{equation*}
H^{k}(\mu )=\left\{ u\in L^{2}(\mathbb{R}^{\Lambda },d\mu ):\partial
^{\alpha }u\in L^{2}(\mathbb{R}^{\Lambda },d\mu )\;\forall \left| \alpha
\right| \leq k\right\} .
\end{equation*}

\begin{proof}
We shall prove that every bounded sequence in $H^{1}(\mu )$ has a convergent
subsequence in $L^{2}(\mathbb{R}^{\Lambda },d\mu )$. Let $\left\{
u_{k}\right\} \subset H^{1}(\mu )=H^{1}(\mathbb{R}^{\Lambda },d\mu )$ be
such that 
\begin{equation*}
\left\| u_{k}\right\| _{H_{\mu }^{1}}\leq \sqrt{M}\text{ \ \ \ for every }k%
\text{ \ \ and some }M>0.
\end{equation*}%
For any $R>0,$ denote by $B(0,R)$ the open ball centered at $0$ with radius $%
R.$ It is clear that $H^{1}(\mathbb{R}^{\Lambda },d\mu )\subset
H^{1}(B(0,R),d\mu )$. Hence $\left\{ u_{k}\right\} $ is a bounded sequence $%
H^{1}(B(0,R),d\mu ).$Moreover 
\begin{eqnarray*}
&&\int_{B(0,R)}u_{k}^{2}dx+\int_{B(0,R)}\left| Du_{k}\right| dx \\
&\leq &C_{\Phi ,R}\left[ \int_{B(0,R)}u_{k}^{2}e^{-\Phi
}dx+\int_{B(0,R)}\left| Du_{k}\right| e^{-\Phi }dx\right] .
\end{eqnarray*}%
This implies that $\left\{ u_{k}\right\} $ is a bounded sequence in $%
H^{1}(B(0,R))$. Now using the standard Sobolev compactness embedding theorem
for bounded domains with nice boundary $\left( \text{see }[3]\right) $, we
get the compactness of the embedding%
\begin{equation*}
H^{1}(B(0,R))\hookrightarrow L^{2}(B(0,R)).
\end{equation*}%
Therefore, one can find a subsequence $\left\{ u_{k_{j}}\right\} $ of $%
\left\{ u_{k}\right\} $ such that $u_{k_{j}}$ converges in $L^{2}(B(0,R)).$%
We shall prove that $\left\{ u_{k_{j}}\right\} $ is Cauchy in $L^{2}(\mathbb{%
R}^{\Lambda },d\mu ).$ Let $\eta >0.$The assumption of the lemma implies
that 
\begin{equation}
\zeta :=\left| \mathbf{\nabla }\Phi \right| ^{2}-(1+\eta )\mathbf{\Delta }%
\Phi
\end{equation}%
is positive in a neighborhood of $\infty $ when $\theta =(1+\eta )^{-1}.$%
\begin{eqnarray}
\int_{\mathbb{R}^{\Lambda }}\left| u_{k_{j}-}u_{k_{l}}\right| ^{2}e^{-\Phi
}dx &\leq &\int_{\left| x\right| <R}\left| u_{k_{j}-}u_{k_{l}}\right|
^{2}e^{-\Phi }+\int_{\left| x\right| \geq R}\frac{\zeta \left|
u_{k_{j}-}u_{k_{l}}\right| ^{2}}{\inf\limits_{\mathbb{R}^{\Lambda
}\;\backslash \;B(0,R)}\zeta }e^{-\Phi }dx \\
&\leq &C_{\Phi }\int_{\left| x\right| <R}\left| u_{k_{j}-}u_{k_{l}}\right|
^{2}+\int_{\left| x\right| \geq R}\frac{\zeta \left|
u_{k_{j}-}u_{k_{l}}\right| ^{2}}{\inf\limits_{\mathbb{R}^{\Lambda
}\;\backslash \;B(0,R)}\zeta }e^{-\Phi }dx.
\end{eqnarray}%
To estimate the last term of the right hand side of this last above
inequality, let $\varepsilon >0$ and choose $R$ large enough so that 
\begin{equation*}
\inf\limits_{\mathbb{R}^{\Lambda }\;\backslash \;B(0,R)}\zeta \geq \frac{%
4M(2+\eta +\eta ^{-1})}{\varepsilon }.
\end{equation*}%
Now introduce the vector fields 
\begin{equation}
X_{j}=\partial _{j}
\end{equation}%
and their formal adjoint in $L^{2}(\mu )$ 
\begin{equation}
X_{j}^{\ast }=-\partial _{j}+\Phi _{x_{j}},
\end{equation}%
one has when $u\in C_{o}^{\infty }(\mathbb{R}^{\Lambda })$ for their sum and
commutator 
\begin{equation}
\left( X_{j}+X_{j}^{\ast }\right) u=\Phi _{x_{j}}u
\end{equation}%
and%
\begin{equation}
\left[ X_{j},X_{j}^{\ast }\right] u=\Phi _{x_{j}x_{j}}u.
\end{equation}%
It is then straightforward to see that 
\begin{equation}
\left( \left[ X_{j},X_{j}^{\ast }\right] u,u\right) _{\mu }=\left\|
X_{j}^{\ast }u\right\| _{\mu }^{2}-\left\| X_{j}u\right\| _{\mu }^{2}
\end{equation}%
\begin{equation}
\left\| \left( X_{j}+X_{j}^{\ast }\right) u\right\| _{\mu }^{2}\leq \left( 1+%
\frac{1}{\eta }\right) \left\| X_{j}u\right\| _{\mu }^{2}+\left( 1+\eta
\right) \left\| X_{j}^{\ast }u\right\| _{\mu }^{2}\;,\;\ \ \forall
\varepsilon >0\;
\end{equation}%
so that a linear combination of these formulae gives for any $\eta >0$ 
\begin{equation}
\left( (\left| \mathbf{\nabla }\Phi \right| ^{2}-(1+\eta )\mathbf{\Delta }%
\Phi )u,u\right) _{\mu }\leq (2+\eta +\eta ^{-1})\left( \left\|
X_{1}u\right\| _{\mu }^{2}+...+\left\| X_{m}u\right\| _{\mu }^{2}\right) .
\end{equation}%
Thus,%
\begin{equation}
\left( (\zeta u,u\right) _{\mu }\leq (2+\eta +\eta ^{-1})\left\| u\right\|
_{H^{1}(\mu )}^{2}
\end{equation}%
Because $C_{o}^{\infty }(\mathbb{R}^{\Lambda })$ is dense in $H^{1}(\mu ),$
this inequality is valid for all $u\in H^{1}(\mu ).$ Now applying $\left(
39\right) $ with $u$ replaced by $u_{k_{j}-}u_{k_{l}},$ $\left( 31\right) $
gives%
\begin{eqnarray*}
\int_{\mathbb{R}^{\Lambda }}\left| u_{k_{j}-}u_{k_{l}}\right| ^{2}e^{-\Phi
}dx &\leq &C_{\Phi }\int_{\left| x\right| <R}\left|
u_{k_{j}-}u_{k_{l}}\right| ^{2}+\frac{(2+\eta +\eta ^{-1})\left\|
u_{k_{j}-}u_{k_{l}}\right\| _{H^{1}(\mu )}^{2}}{4M(2+\eta +\eta ^{-1})}%
\varepsilon \\
&\leq &C_{\Phi }\int_{\left| x\right| <R}\left| u_{k_{j}-}u_{k_{l}}\right|
^{2}+\varepsilon
\end{eqnarray*}%
The result follows from the convergence of the subsequence $\left\{
u_{k_{j}}\right\} $ in $L^{2}(B(0,R)).$
\end{proof}

The Lemma above indicates the direction towards the assumptions needed for
the Hamiltonian $\Phi =\Phi _{\Lambda }.$\newline
\textbf{Assumptions on }$\Phi .$

Recall that $\Lambda $\ is a finite domain in $\mathbb{Z}^{d}$ $.$ We shall
assume that $\Phi (x)\in C^{\infty }(\mathbb{R}^{\Lambda })$ satisfying:

\begin{tabular}{l}
1.$\;\lim\limits_{\left| x\right| \rightarrow \infty }\left| \mathbf{\nabla }%
\Phi (x)\right| =\infty $ \\ 
2. For some $M,$ any $\partial ^{\alpha }\Phi $ with $\left| \alpha \right|
=M$ \ is bounded on $\mathbb{R}^{\Lambda }.$ \\ 
3. For $\left| \alpha \right| \geq 1,$ there are constants $C_{\alpha }$
such that $\left| \partial ^{\alpha }\Phi (x)\right| \leq C_{\alpha }\left(
1+\left| \mathbf{\nabla }\Phi (x)\right| ^{2}\right) ^{1/2}$ \\ 
4. $\mathbf{Hess}\Phi \geq \delta $ for some $0<\delta \leq 1$%
\end{tabular}

\section{Preliminary Results on Hilbert Space Methods For Elliptic PDE}

A bilinear form with domain $H,$ a complex Hilbert space, is a
complex-valued function $a$ defined on $H\times H$ which is such that $%
a(u,v) $ is linear in $u$ and conjugate linear in $v.$The inner product $%
\left( \cdot ,\cdot \right) _{H}$ on $H$ is clearly a bilinear form; we
shall denote it by $1\left( \cdot ,\cdot \right) .$The form $a+\lambda 1$
will simply be denoted by $a+\lambda $%
\begin{equation*}
\left( a+\lambda \right) (u,v)=a(u,v)+\lambda (u,v)_{H}.
\end{equation*}%
The adjoint $a^{\ast }$ of $a$ is defined by 
\begin{equation*}
a^{\ast }(u,v)=\overline{a(v,u)}
\end{equation*}%
and $a$ is said to be symmetric if $a\equiv a^{\ast },$ i.e. for all $u,v\in
H$%
\begin{equation*}
a^{\ast }(u,v)=\overline{a(v,u)}=a(u,v).
\end{equation*}%
A bilinear form is said to be bounded on $H\times H$ if there exists a
constant $M>0$ such that 
\begin{equation*}
\left| a(u,v)\right| \leq M\left\| u\right\| _{H}\left\| v\right\| _{H}\text{
\ \ \ \ for all }u,v\in H.
\end{equation*}%
A bilinear form $a$ is said to be coercive on $H$ if there exists a positive
constant $m>0$ such that%
\begin{equation*}
\left| a(u,u)\right| \geq m\left\| u\right\| _{H}^{2}\text{\ \ \ \ \ for all 
}u,v\in H.
\end{equation*}

We shall say that a Banach space $W$ is continuously embedded in a Banach
space $X$ if there is a bounded operator $E:W\rightarrow X$ which is
one-to-one. We call $E$ an embedding operator. We shall say that $W$ is
densely embedded in $X$ if $R(E),$ the range of $E$ is dense in $X;$ and we
shall write 
\begin{equation*}
W\hookrightarrow _{ds}^{E}X.
\end{equation*}%
If $X$ is a Banach Space, the set of all linear conjugate functionals on $X$
shall be denoted by $X^{\ast }$ and is called the conjugate space of $%
X^{\ast }.$

Suppose $X,Y,W,Z$ are Banach spaces such that%
\begin{equation*}
W\hookrightarrow _{ds}^{E}X\text{ \ \ \ \ \ \ \ \ and \ \ \ \ \ \ \ \ }%
Y\hookrightarrow _{ds}^{F}Z^{\ast }.
\end{equation*}%
Let $a(w,z)$ be a bounded bilinear form on $W\times Z.$ We can define two
linear operators connected with $a(w,z).$ The first which we shall denote by 
$A,$ is an operator from $X$ to $Y.$ We say that $x\in D(A),$ the domain of $%
A$ and $Ax=y$ if $x\in R(E),$ $y\in Y$ and 
\begin{equation*}
a(E^{-1}x,z)=Fy(z),\ \ \ \ \ \ \ \text{for all }z\in Z.
\end{equation*}%
Since $R(F)$ is dense in $Z^{\ast },$ the operator $A$ is well defined. We
call $A$ the operator associated with the bilinear form $a(u,v).$\newline
The second operator, which we denote by $\hat{A}$, is from $W$ to $Z^{\ast
}. $ We define it as follows. Fix $w\in W,$ $a(w,\cdot )\in Z^{\ast },$ it
is bounded because the bilinear form $a$ is bounded. We define $\hat{A}w$ to
be $a(w,\cdot ).$ $\hat{A}$ is clearly well defined and will be called the
extended linear operator associated with the bilinear form $a(u,v).$ It can
be shown that $A$ and $\hat{A}$ are related in the following way:%
\begin{equation*}
A=F^{-1}\hat{A}E^{-1}
\end{equation*}%
The fundamental tool to investigate the operator $\hat{A}$ is the
Lax-Milgram theorem

\begin{theorem}[Lax Milgram]
\textit{Let }$a$\textit{\ be a bounded coercive form on a Hilbert space }$%
H_{o}$\textit{\ with bounds }$m$ \textit{and }$M$\textit{\ as above. Then
for any }$F\in H_{o}^{\ast },$\textit{\ the adjoint of }$H_{o},$\textit{\
there exists an }$u$\textit{\ }$\in H_{o}$\textit{\ such that }%
\begin{equation*}
a(u,v)=\left\langle F,v\right\rangle \text{ \ \ \ \ \ \ \ for all }v\in H_{o}
\end{equation*}%
\textit{The map }$\hat{A}:u\mapsto F$\textit{\ defined above is a linear
bijection of }$H_{o}$\textit{\ onto }$H_{o}^{\ast }$\textit{\ and }%
\begin{equation*}
m\leq \left\Vert \hat{A}\right\Vert \leq M,\ \ \ \ \ \ \ \ \ \ M^{-1}\leq
\left\Vert \hat{A}^{-1}\right\Vert \leq m^{-1}.
\end{equation*}
\end{theorem}

\begin{proof}
\textbf{\ }(see [3]).
\end{proof}

\begin{corollary}
\textit{For any choice of }$F\in H_{o}^{\ast }$\textit{\ there is a unique
vector }$u\in H_{o}$\textit{\ satisfying }%
\begin{equation*}
\left( u,v\right) _{H_{0}}=F(v)\text{ \ \ \ \ \ \ \ for all }v\in H_{o};
\end{equation*}%
\textit{moreover, the isomorphism }$\hat{A}^{-1}$\textit{\ from }$%
H_{o}^{\ast }$\textit{\ onto }$H_{o}$\textit{\ defined by }$\hat{A}^{-1}F=u$%
\textit{\ verifies }%
\begin{equation*}
\left\Vert \hat{A}^{-1}F\right\Vert _{H_{o}}=\left\Vert F\right\Vert
_{H_{o}^{\ast }}
\end{equation*}
\end{corollary}

Next, we apply the Lax-Milgram theorem to the situation where the Hilbert
space $H_{o}$ is continuously and densely embedded in another Hilbert $H.$

\begin{lemma}
\textit{If }$H$\textit{\ is a Hilbert space and }$W$\textit{\ is a Banach
space continuously and densely embedded in }$H$\textit{\ with embedding
operator }$E,$\textit{\ then }$H$\textit{\ can be continuously and densely
embedded in }$W^{\ast }$\textit{\ with embedding operator }$F$\textit{\
satisfying }%
\begin{equation*}
\left( x,Ew\right) _{H}=Fx(w),\text{ \ \ \ \ }x\in H,\text{ \ \ and \ \ }%
w\in W.
\end{equation*}
\end{lemma}

\begin{proof}
For each $x\in H,$ the function $x^{\ast }:w\longmapsto \left( x,Ew\right)
_{H}$ is a conjugate linear functional on $W$ and 
\begin{equation*}
\left\vert x^{\ast }(x)\right\vert \leq \left\Vert x\right\Vert
_{H}\left\Vert E\right\Vert \left\Vert w\right\Vert _{W}.
\end{equation*}%
Hence $x^{\ast }\in W^{\ast }.$ Define the operator $F$ from $H$ to $W^{\ast
}$ by $Fx=x^{\ast }.$ Clearly, $F$ is linear and bounded. It is also
one-to-one since $R(E)$ is dense in $H.$ Finally, suppose $x^{\ast }(w)=0$
for all $x^{\ast }\in R(F).$ Then $\left( x,Ew\right) _{H}=0$ for all $x\in
H.$ Thus $Ew=0$ and consequently $w=0.$This shows that $R(F)$ is dense
\end{proof}

Now let the Hilbert space $H_{o}$ be continuously and densely embedded into
another Hilbert space $H$ with embedding operator $E.$ By the lemma above, $%
H $ can be continuously and densely embedded in $H_{o}^{\ast }$ with
embedding operator $F.$We obtain the scheme 
\begin{equation*}
H_{o}\hookrightarrow _{ds}^{E}H\hookrightarrow _{ds}^{F}H_{o}^{\ast }
\end{equation*}%
which is referred to by saying that $\left( H_{o},H,H_{o}^{\ast }\right) $
is a Hilbert triplet. Notice also that if the embedding $E$ is compact, then
so is the embedding%
\begin{equation*}
H_{o}\hookrightarrow ^{FE}H_{o}^{\ast }.
\end{equation*}%
Returning to the bilinear form on $H_{o}$, we weaken the notion of
coerciveness as follows: We say that a bilinear form $a(u,v)$ on $H_{o}$ is
coercive relative to $H$, if there exists some $\lambda >0$ such that $%
a_{\lambda }(u,v)=a(u,v)+\lambda \left( u,v\right) _{H}$ is coercive, i.e.%
\begin{equation*}
a(u,u)+\lambda \left\| u\right\| _{H}^{2}\geq \alpha _{o}\left\| u\right\|
_{H_{o}}^{2}\ \ \ \ \ \ \text{for }u\in H_{o}\text{ and some }\alpha _{o}>0.
\end{equation*}%
If this last inequality above holds, then by Lax-Milgram, the extended
linear operator $\hat{A}_{\lambda }$ associated with the bilinear form $%
a_{\lambda }(u,v)$ has a bounded inverse $\hat{A}_{\lambda }^{-1}:$ $%
H_{o}^{\ast }\rightarrow H_{o},$ moreover $\hat{A}_{\lambda }u=\hat{A}%
u+\lambda \hat{B}u,$ where $\hat{A}$ is the extended operator associated
with the bilinear form $a(u,v)$ and $\hat{B}$ the extended operator
associated with the inner product $\left( u,v\right) _{H}.$

Now Let $q\in H_{o}^{\ast }$ and consider the equation%
\begin{equation}
u\in H_{o},\ \ \ \ \ \hat{A}u=q
\end{equation}%
$\left( 1.8\right) $ can now be written as 
\begin{equation}
u\in H_{o},\ \ \ \ \ u-\lambda \hat{A}_{\lambda }^{-1}\hat{B}u=z
\end{equation}%
with $z=\hat{A}_{\lambda }^{-1}q.$We now claim that the compactness of the
embedding $E$ implies that of the operator $\hat{A}_{\lambda }^{-1}\hat{B}%
:H_{o}\rightarrow H_{o}$ is compact. Indeed this follows from the fact that $%
\hat{B}$ is bounded and $\hat{A}_{\lambda }^{-1}:$ $H_{o}^{\ast }\rightarrow
H_{o}$ is compact. By the Fredholm alternative (see theorem 3 below), $%
\left( 1.9\right) $ is uniquely solvable for any choice of $z\in H_{o}$ if
and only if $u=0$ is the unique vector of $H_{o}$ satisfying $\ u-\lambda 
\hat{A}_{\lambda }^{-1}\hat{B}u=0.$ When this is the case, the linear
operator $z\longmapsto u$ defined by $\left( 1.9\right) $ is bounded from $%
H_{o}$ to $H_{o}$. Summing up, we have the following theorem

\begin{theorem}
\textit{Let }$\left( H_{o},H,H_{o}^{\ast }\right) $\textit{\ be a Hilbert
triplet with }$H_{o}$\textit{\ compactly embedded in }$H,$\textit{\ let }$%
a(u,v)$\textit{\ be a bounded bilinear form on }$H_{o}$\textit{\ coercive
relative to }$H.$\textit{\ Then }%
\begin{equation*}
u\in H_{o},\ \ \ \ \ a(u,v)=q(v)\ \ \ \ \ \ \ \text{for }v\in H_{o}
\end{equation*}%
\textit{admits a unique solution }$u$\textit{\ for any choice of }$q\in
H_{o}^{\ast }$\textit{\ if and only if it admits a unique solution }$u=0$%
\textit{\ for }$q=0$\textit{\ in which case the solution }$u$\textit{\
satisfies }%
\begin{equation*}
\left\| u\right\| _{H_{o}}\leq C\left\| q\right\| _{H_{o}}
\end{equation*}%
\textit{with }$C$\textit{\ dependent only on }$\hat{A}.$
\end{theorem}

\begin{theorem}[Fredholm Alternative]
\textit{Let }$T$\textit{\ be a compact linear operator on a Hilbert space }$%
V $\textit{\ and consider the equations }%
\begin{equation}
u\in V,\;\;\;\;u-Tu=f
\end{equation}%
\begin{equation}
v\in V,\;\;\;\;v^{\ast }-T^{\ast }v^{\ast }=g
\end{equation}%
\textit{where }$T^{\ast }$\textit{\ the adjoint operator of }$T.$\textit{\
Then the following alternative holds:\newline
(i) either there exists a unique solution of }$\left( 42\right) $\textit{\
and }$\left( 43\right) $\textit{\ for any }$f$\textit{\ and }$g$\textit{\ in 
}$V,$\textit{\ or}\newline
\textit{(ii) the homogeneous equation}%
\begin{equation*}
u-Tu=0
\end{equation*}%
\textit{has nontrivial solutions. In that case the dimension of the null
space of }$I-T$\textit{\ \ is finite and equals the dimension\ of the null
space }$\mathcal{N}^{\ast }$\textit{\ of }$I-T^{\ast }$\textit{Furthermore }$%
\left( 42\right) $ \textit{and} $\left( 43\right) $\textit{\ have solutions
(not unique ) if and only if }%
\begin{equation*}
\left\langle f,v^{\ast }\right\rangle =0,\;\;\;\;\forall v\in \mathcal{N}%
^{\ast }\;
\end{equation*}%
\textit{and \ }%
\begin{equation*}
\left\langle g,v\right\rangle =0,\;\;\;\;\forall v\in \mathcal{N}\;
\end{equation*}%
$\mathcal{N}\;$being the null space of $I-T$
\end{theorem}

\begin{proof}
$.$see Yosida [17],$\left( X-\S 5\right) $
\end{proof}

\begin{remark}
Assumption 2 implies that $\Phi $ is a slowly increasing function. This
assumption is made to rule out any possibility of exponential growth for $%
\Phi .$
\end{remark}

\section{Solvability and Regularity of The Basic Equation}

\begin{theorem}
\textit{Let }$\Lambda $\textit{\ be a finite domain in }$\mathbb{Z}^{d}$%
\textit{.\ If }$\Phi $ \textit{satisfies assumptions 1-4 above,} \textit{%
then for any }$C^{\infty }-$\textit{function }$g$\textit{\ satisfying }%
\begin{equation}
\left| D^{\alpha }g\right| \leq C_{\alpha }(1+Z_{\Phi })^{q_{\alpha }}
\end{equation}%
\textit{where}%
\begin{equation*}
Z_{\Phi }=\frac{\left| \mathbf{\nabla }\Phi \right| }{2},
\end{equation*}%
$\alpha \in \mathbb{N}^{\left| \Lambda \right| }$\textit{\ with some }$%
C_{\alpha }$\textit{\ and some }$q_{\alpha }>0,$\textit{\ there exists a
unique }$C^{\infty }-$\textit{function }$u$\textit{\ solution of \ }%
\begin{equation}
\left\{ 
\begin{tabular}{l}
$A_{\Phi }^{(0)}v=g-\left\langle g\right\rangle _{_{L^{2}(\mu )}}$ \\ 
$\left\langle v\right\rangle _{L^{2}(\mu )}=0.$%
\end{tabular}%
\right.
\end{equation}
\end{theorem}

\begin{proof}
(\textbf{Existence}) We shall work in the unweighted space $L^{2}(\mathbb{R}%
^{\Lambda })$ and with the Witten-Laplacians ensuing after the unitary
transformation.

Under the unitary transformation,%
\begin{equation*}
A_{\Phi }^{(0)}v=g-\left\langle g\right\rangle _{_{L^{2}(\mu )}\text{\ \ \ \
\ }}\text{ in }\mathbb{R}^{\Lambda }
\end{equation*}%
is equivalent to%
\begin{equation*}
\mathbf{W}_{\Phi }^{\left( 0\right) }u=q\;\;\;\;\ \ \ \ \ \ \ \ \ \ \ \ \ 
\text{in }\mathbb{R}^{\Lambda }
\end{equation*}%
where 
\begin{equation*}
u=e^{-\Phi /2}v\;\;\ \;\text{and \ \ \ \ }q=e^{-\Phi /2}(g-\left\langle
g\right\rangle _{_{L^{2}(\mu )}})\in L^{2}(\mathbb{R}^{\Lambda }).
\end{equation*}%
Let%
\begin{equation*}
B_{\Phi }^{k}(\mathbb{R}^{\Lambda })=\left\{ u\in L^{2}(\mathbb{R}^{\Lambda
}):Z_{\Phi }^{l}\partial ^{\alpha }u\in L^{2}(\mathbb{R}^{\Lambda
})\;\forall \;l+\left| \alpha \right| \leq k\right\}
\end{equation*}%
(here $\partial ^{\alpha }u$ is taken in the distributional sense in $%
\mathbb{R}^{\Lambda }).$

Denote by $B_{o,\Phi }^{1}(\mathbb{R}^{\Lambda })$ be the closure of $%
C_{o}^{\infty }(\mathbb{R}^{\Lambda })$ in $B_{\Phi }^{1}(\mathbb{R}%
^{\Lambda }),$ and let $\mathbf{b}$ be the bilinear form on $B_{o,\Phi }^{1}(%
\mathbb{R}^{\Lambda })$ defined by 
\begin{equation*}
\mathbf{b}:B_{o,\Phi }^{1}(\mathbb{R}^{\Lambda })\times B_{o,\Phi }^{1}(%
\mathbb{R}^{\Lambda })\rightarrow \mathbb{R}
\end{equation*}%
with 
\begin{equation*}
\mathbf{b}(u,w)=\int_{\mathbb{R}^{\Lambda }}Du\cdot Dwdx+\int_{\mathbb{R}%
^{\Lambda }}\left( \dfrac{\left| \mathbf{\nabla }\Phi \right| ^{2}}{4}-%
\dfrac{\mathbf{\Delta }\Phi }{2}\right) uwdx.
\end{equation*}%
Because we have in mind to apply theorems 6 and 7 above, we need to check
boundedness and coerciveness of $\mathbf{b.}$

\textit{Boundedness:} After observing that 
\begin{eqnarray*}
\mathbf{\Delta }\Phi &\leq &C(1+\left| \mathbf{\nabla }\Phi \right|
^{2})^{1/2} \\
&\leq &C(1+\left| \mathbf{\nabla }\Phi \right| ^{2}),
\end{eqnarray*}%
it then follows immediately from Cauchy-Schwartz inequality that 
\begin{equation*}
\left| \mathbf{b}(u,w)\right| \leq \alpha _{o}\left\| u\right\| _{B_{\Phi
}^{1}(\mathbb{R}^{\Lambda })}\left\| w\right\| _{B_{\Phi }^{1}(\mathbb{R}%
^{\Lambda })}
\end{equation*}%
for some constant $\alpha _{o}>0.$

\textit{Coerciveness:}%
\begin{equation*}
\int_{\mathbb{R}^{\Lambda }}\left| Du\right| ^{2}dx=\mathbf{b}(u,u)-\int_{%
\mathbb{R}^{\Lambda }}\left( \dfrac{\left| \mathbf{\nabla }\Phi \right| ^{2}%
}{4}-\dfrac{\mathbf{\Delta }\Phi }{2}\right) \left| u\right| ^{2}dx
\end{equation*}%
\begin{eqnarray*}
\int_{\mathbb{R}^{\Lambda }}\left| Du\right| ^{2}dx+\int_{\mathbb{R}%
^{\Lambda }}\left| Z_{\Phi }u\right| ^{2}dx &=&\mathbf{b}(u,u)+\int_{\mathbb{%
R}^{\Lambda }}\dfrac{\mathbf{\Delta }\Phi }{2}\left| u\right| ^{2}dx \\
&\leq &\mathbf{b}(u,u)+\varepsilon \int_{\mathbb{R}^{\Lambda }}\dfrac{\left( 
\mathbf{\Delta }\Phi \right) ^{2}}{4}\left| u\right| ^{2}dx+\frac{1}{%
4\varepsilon }\int_{\mathbb{R}^{\Lambda }}\left| u\right| ^{2}dx \\
&\leq &\mathbf{b}(u,u)+C\varepsilon \int_{\mathbb{R}^{\Lambda }}\left|
Z_{\Phi }u\right| ^{2}dx+\left( C\varepsilon +\frac{1}{4\varepsilon }\right)
\int_{\mathbb{R}^{\Lambda }}\left| u\right| ^{2}dx
\end{eqnarray*}%
choosing $\varepsilon $ such that $C\varepsilon <1$ and adding $\int_{%
\mathbb{R}^{\Lambda }}\left| u\right| ^{2}dx$ on both side of this above
inequality, we immediately get 
\begin{equation}
\delta \left\| u\right\| _{B_{\Phi }^{1}(\mathbb{R}^{\Lambda })}^{2}\leq 
\mathbf{b(}u,u\mathbf{)+\gamma }\left\| u\right\| _{L^{2}(\mathbb{R}%
^{\Lambda })}^{2}
\end{equation}%
for some positive constants $\delta $ and $\gamma .$

This shows that the bilinear form $\mathbf{b}(u,v)$ is bounded and coercive
relative to $L^{2}(\mathbb{R}^{\Lambda }).$

Observe that $B_{o,\Phi }^{1}(\mathbb{R}^{\Lambda })$ is densely embedded
into $L^{2}(\mathbb{R}^{\Lambda }).$ Now considering the Hilbert triplet 
\begin{equation}
\left( B_{o,\Phi }^{1}(\mathbb{R}^{\Lambda }),L^{2}(\mathbb{R}^{\Lambda
}),B_{o,\Phi }^{-1}(\mathbb{R}^{\Lambda })\right) ,
\end{equation}%
where $B_{o,\Phi }^{-1}(\mathbb{R}^{\Lambda })$ denote the conjugate space
of $B_{o,\Phi }^{1}(\mathbb{R}^{\Lambda }).$

We need to check that the embedding 
\begin{equation*}
B_{o,\Phi }^{1}(\mathbb{R}^{\Lambda })\hookrightarrow L^{2}(\mathbb{R}%
^{\Lambda })
\end{equation*}%
is compact. This follows from Lemma 1 by simply observing that 
\begin{equation*}
B_{o,\Phi }^{1}(\mathbb{R}^{\Lambda })\subset U_{\Phi }^{-1}\left( H^{1}(\mu
)\right)
\end{equation*}%
and the fact that $U_{\Phi }$ is a unitary operator.

Let $\mathbf{B}_{\gamma }$ be the bilinear form in $B_{o,\Phi }^{1}(\mathbb{R%
}^{\Lambda })$ defined by\ 
\begin{equation*}
\mathbf{B}_{\gamma }(u,w)=\mathbf{b(}u,w\mathbf{)+\gamma }\left\langle
u,w\right\rangle _{L^{2}(\mathbb{R}^{\Lambda })}
\end{equation*}%
and%
\begin{equation*}
\hat{A}_{\gamma }:B_{o,\Phi }^{1}(\mathbb{R}^{\Lambda })\rightarrow
B_{o,\Phi }^{-1}(\mathbb{R}^{\Lambda })
\end{equation*}%
be the extended linear operator associated with the bilinear form $\mathbf{B}%
_{\gamma }(u,w).$We have 
\begin{equation}
\hat{A}_{\gamma }u=\hat{A}u+\gamma \hat{B}u,
\end{equation}%
where $\hat{A}$ and $\hat{B}$ are the bounded bilinear forms associated with 
$\mathbf{b}$ and $\left( \cdot ,\cdot \right) _{L^{2}}$ respectively.

Note that the equation 
\begin{equation*}
u\in B_{o,\Phi }^{1}(\mathbb{R}^{\Lambda })\ \ \ \ \ \ \ \ \hat{A}u=q
\end{equation*}%
is the variational interpretation of the equation 
\begin{equation*}
\mathbf{W}_{\Phi }^{\left( 0\right) }u=q\;\;\;\;\ \ \ \ \ \ \ \ \ \ \text{in 
}\mathbb{R}^{\Lambda }.
\end{equation*}%
By theorem 1 (Lax-Milgram), the boundedness of $\mathbf{B}_{\gamma }$ and
the coercivity condition 
\begin{equation*}
\mathbf{B}_{\gamma }(u,u)\geq \delta \left\| u\right\| _{B^{1}(\mathbb{R}%
^{\Lambda })}^{2}\;\;\;\;\forall u\in B_{o,\Phi }^{1}(\mathbb{R}^{\Lambda })
\end{equation*}%
guarantee that $A_{\gamma }$ has a bounded inverse 
\begin{equation*}
\hat{A}_{\gamma }^{-1}:B_{o,\Phi }^{-1}(\mathbb{R}^{\Lambda })\rightarrow
B_{o,\Phi }^{1}(\mathbb{R}^{\Lambda }).
\end{equation*}%
Now using the fact that 
\begin{equation*}
\hat{A}_{\gamma }u=\hat{A}u+\gamma \hat{B}u,
\end{equation*}%
we can write the equation 
\begin{equation*}
u\in B_{o,\Phi }^{1}(\mathbb{R}^{\Lambda })\ \ \ \ \ \ \ \ \hat{A}u=q
\end{equation*}%
as 
\begin{equation}
u\in B_{o,\Phi }^{1}(\mathbb{R}^{\Lambda }),\;\;\;\;\;\;u-\gamma \hat{A}%
_{\gamma }^{-1}\hat{B}u=z
\end{equation}%
where%
\begin{equation}
z=\hat{A}_{\gamma }^{-1}q.
\end{equation}%
As in the preliminary, because the injection 
\begin{equation*}
B_{o,\Phi }^{1}(\mathbb{R}^{\Lambda })\hookrightarrow L^{2}(\mathbb{R}%
^{\Lambda })
\end{equation*}%
is compact, the operator $\gamma \hat{A}_{\gamma }^{-1}\hat{B}:B_{o,\Phi
}^{1}(\mathbb{R}^{\Lambda })\rightarrow B_{o,\Phi }^{1}(\mathbb{R}^{\Lambda
})$ is compact. Moreover, the boundedness of $\gamma \hat{A}_{\gamma }^{-1}%
\hat{B}$ implies that 
\begin{eqnarray}
\left( \gamma \hat{A}_{\gamma }^{-1}\hat{B}\right) ^{\ast } &=&\left( \gamma
\left( \hat{B}_{\gamma }^{-1}\hat{A}_{\gamma }\right) ^{\ast }\right) ^{-1}
\\
&=&\gamma \left( \hat{A}_{\gamma }^{\ast }\left( \hat{B}^{-1}\right) ^{\ast
}\right) ^{-1} \\
&=&\gamma \left( \hat{A}_{\gamma }^{\ast }\left( \hat{B}^{\ast }\right)
^{-1}\right) ^{-1} \\
&=&\gamma \hat{A}_{\gamma }^{-1}\hat{B}.
\end{eqnarray}%
Let us also point out that the self-adjointness of $\hat{A}_{\gamma }$ and $%
\hat{B}$ follow from the fact that they are both associated with symmetric
bilinear forms.\newline
Now observe that 
\begin{equation}
\ker (I-\gamma \hat{A}_{\gamma }^{-1}\hat{B})\subset \ker \hat{A}.
\end{equation}%
We now claim that 
\begin{equation}
\ker \hat{A}=\left\{ \delta e^{-\Phi /2},\delta \in \mathbb{R}\right\} .
\end{equation}%
Indeed if $\hat{A}u=0$ , then $\mathbf{b}(u,u)=0.$ Hence 
\begin{equation*}
\left\| \left( \partial _{x}+\dfrac{\mathbf{\nabla }\Phi }{2}\right)
u\right\| _{L^{2}}^{2}=0
\end{equation*}%
which would imply that $u$ is a solution of the equation%
\begin{equation*}
\left( \partial _{x}+\dfrac{\mathbf{\nabla }\Phi }{2}\right) u=0.
\end{equation*}%
One can then easily see $u$ must be a constant multiple of $e^{-\Phi /2}.$We
have in mind to apply the second part of Theorem 7 (Fredholm alternative).
This brings us to check orthogonality of $q$ with $\ker (I-\gamma \hat{A}%
_{\gamma }^{-1}\hat{B}).$ Let $\delta \in \mathbb{R},$%
\begin{eqnarray}
\left\langle \delta e^{-\Phi /2},q\right\rangle _{L^{2}(\mathbb{R}^{\Lambda
})} &=&\int_{\mathbb{R}^{\Lambda }}\delta e^{-\Phi /2}e^{-\Phi
/2}(g-\left\langle g\right\rangle _{_{L^{2}(\mu )}}) \\
&=&\delta \left( \left\langle g\right\rangle _{_{L^{2}(\mu )}}-\left\langle
g\right\rangle _{_{L^{2}(\mu )}}\right) =0.
\end{eqnarray}%
Hence using part (ii) of theorem 3, we conclude that the equation 
\begin{equation}
\hat{A}u=q
\end{equation}%
is solvable therefore%
\begin{equation}
A_{\Phi }^{(0)}v=g-\left\langle g\right\rangle _{_{L^{2}(\mu )}}
\end{equation}%
is solvable in the weak\ sense. To complete the proof of theorem 4, we need
to prove that the $L^{2}-$solution constructed above is a classical solution.
\end{proof}

\textbf{Regularity: }Next, we shall prove that the weak solutions
constructed above are actually classical solutions. The proof is based on
the method of difference quotient.

\begin{theorem}[$B^{k}$-regularity]
\textit{Given }$q\in B_{\Phi }^{k-1}(\mathbb{R}^{\Lambda })$ \textit{for }$%
k=0,1,2,...,$ \textit{a solution }$u\in B_{o,\Phi }^{1}(\mathbb{R}^{\Lambda
})$\textit{\ of }%
\begin{equation}
\hat{A}u=q
\end{equation}%
\textit{is an element of }$B_{\Phi }^{k+1}(\mathbb{R}^{\Lambda })$ \textit{%
and we have the estimate}%
\begin{equation}
\left\Vert u\right\Vert _{B_{\Phi }^{k+1}(\mathbb{R}^{\Lambda })}\leq C\left[
\left\Vert \hat{A}u\right\Vert _{B_{\Phi }^{k-1}(\mathbb{R}^{\Lambda
}))}+\left\Vert u\right\Vert _{B_{\Phi }^{k}(\mathbb{R}^{\Lambda })}\right]
\end{equation}%
\textit{for all }$u\in B_{\Phi }^{k+1}(\mathbb{R}^{\Lambda }).$
\end{theorem}

\begin{proof}
We first establish the result when $k=0.$ We have%
\begin{eqnarray}
\left( \dfrac{\mathbf{\Delta }\Phi }{2}u\;,u\right) _{L^{2}} &\leq &\left\| 
\dfrac{\mathbf{\Delta }\Phi }{2}u\right\| _{L^{2}(\mathbb{R}^{\Lambda
})}\left\| u\right\| _{L^{2}(\mathbb{R}^{\Lambda })} \\
&\leq &C\left\| u\right\| _{B_{\Phi }^{1}(\mathbb{R}^{\Lambda })}\left\|
u\right\| _{L^{2}(\mathbb{R}^{\Lambda })} \\
&\leq &\varepsilon C\left\| u\right\| _{B_{\Phi }^{1}(\mathbb{R}^{\Lambda
})}^{2}+\frac{C}{4\varepsilon }\left\| u\right\| _{L^{2}(\mathbb{R}^{\Lambda
})}^{2}.
\end{eqnarray}%
Thus, for $u\in B_{o,\Phi }^{1}(\mathbb{R}^{\Lambda }),$%
\begin{eqnarray*}
\left\langle \hat{A}u,u\right\rangle &=&\left\| Du\right\| _{L^{2}(\mathbb{R}%
^{\Lambda })}^{2}+\left( Z_{\Phi }^{2}u\;,u\right) _{L^{2}}-\left( \dfrac{%
\mathbf{\Delta }\Phi }{2}u\;,u\right) _{L^{2}} \\
&\geq &\left\| Du\right\| _{L^{2}(\mathbb{R}^{\Lambda })}^{2}+\left\|
Z_{\Phi }u\right\| _{L^{2}(\mathbb{R}^{\Lambda })}^{2}-\varepsilon C\left\|
u\right\| _{B_{\Phi }^{1}(\mathbb{R}^{\Lambda })}^{2}-\frac{C}{4\varepsilon }%
\left\| u\right\| _{L^{2}(\mathbb{R}^{\Lambda })}^{2}.
\end{eqnarray*}%
Choosing $\varepsilon $ such that $\varepsilon C<1,$ we get%
\begin{equation*}
\left\langle \hat{A}u,u\right\rangle \geq C\left\| u\right\| _{B_{\Phi }^{1}(%
\mathbb{R}^{\Lambda })}^{2}-C\left\| u\right\| _{L^{2}(\mathbb{R}^{\Lambda
})}^{2}
\end{equation*}%
Hence 
\begin{eqnarray*}
\left\| u\right\| _{B_{\Phi }^{1}(\mathbb{R}^{\Lambda })}^{2} &\leq
&C\left\langle \hat{A}u,u\right\rangle +C\left\| u\right\| _{L^{2}(\mathbb{R}%
^{\Lambda })}^{2} \\
&\leq &C\left\| \hat{A}u\right\| _{B_{\Phi }^{-1}(\mathbb{R}^{\Lambda
})}\left\| u\right\| _{B_{\Phi }^{1}(\mathbb{R}^{\Lambda })}+C\left\|
u\right\| _{L^{2}(\mathbb{R}^{\Lambda })}^{2} \\
&\leq &\frac{C}{4\varepsilon }\left\| \hat{A}u\right\| _{B_{\Phi }^{-1}(%
\mathbb{R}^{\Lambda })}^{2}+C\varepsilon \left\| u\right\| _{B_{\Phi }^{1}(%
\mathbb{R}^{\Lambda })}^{2}+C\left\| u\right\| _{L^{2}(\mathbb{R}^{\Lambda
})}^{2}.
\end{eqnarray*}%
Again choosing $\varepsilon $ appropriately, $\left( \varepsilon C<1\right) $
we finally get 
\begin{equation*}
\left\| u\right\| _{B_{\Phi }^{1}(\mathbb{R}^{\Lambda })}^{2}\leq C\left\| 
\hat{A}u\right\| _{B_{\Phi }^{-1}(\mathbb{R}^{\Lambda })}^{2}+C\left\|
u\right\| _{B_{\Phi }^{0}(\mathbb{R}^{\Lambda })}^{2}.
\end{equation*}%
Now assume that for $u\in B_{o,\Phi }^{1}(\mathbb{R}^{\Lambda }),$ $\hat{A}%
u=q\in B_{\Phi }^{k-1}(\mathbb{R}^{\Lambda })$ implies $u\in B_{\Phi }^{k+1}(%
\mathbb{R}^{\Lambda })$ and that 
\begin{equation}
\left\| u\right\| _{B_{\Phi }^{k+1}(\mathbb{R}^{\Lambda })}\leq C\left[
\left\| \hat{A}u\right\| _{B_{\Phi }^{k-1}(\mathbb{R}^{\Lambda }))}+\left\|
u\right\| _{B_{\Phi }^{k}(\mathbb{R}^{\Lambda })}\right] .
\end{equation}%
Suppose now that $u\in B_{o,\Phi }^{1}(\mathbb{R}^{\Lambda })$, $\hat{A}u\in
B_{\Phi }^{k}(\mathbb{R}^{\Lambda }).$ So we know that $u\in B_{\Phi }^{k+1}(%
\mathbb{R}^{\Lambda })$ and we want to establish that $u\in B_{\Phi }^{k+2}(%
\mathbb{R}^{\Lambda }).$

Because 
\begin{equation*}
D_{i}^{h}u=\dfrac{u(x+he_{i})-u(x)}{h}\in B_{\Phi }^{k+1}(\mathbb{R}%
^{\Lambda }),
\end{equation*}%
replacing $u$ by $D_{i}^{h}u$ in inequality $\left( 66\right) $ we get 
\begin{eqnarray*}
\left\| D_{i}^{h}u\right\| _{B_{\Phi }^{k+1}(\mathbb{R}^{\Lambda })} &\leq &C%
\left[ \left\| \hat{A}D_{i}^{h}u\right\| _{B_{\Phi }^{k-1}(\mathbb{R}%
^{\Lambda })}+\left\| D_{i}^{h}u\right\| _{B_{\Phi }^{k}(\mathbb{R}^{\Lambda
})}\right] \\
&\leq &C\left[ \left\| D_{i}^{-h}\hat{A}u\right\| _{B_{\Phi }^{k-1}(\mathbb{R%
}^{\Lambda })}+\left\| uD_{i}^{h}X_{\Phi }\right\| _{B_{\Phi }^{k-1}(\mathbb{%
R}^{\Lambda }))}+\left\| D_{i}^{h}u\right\| _{B_{\Phi }^{k}(\mathbb{R}%
^{\Lambda })}\right]
\end{eqnarray*}%
where 
\begin{equation*}
X_{\Phi }:=\dfrac{\left| \mathbf{\nabla }\Phi \right| ^{2}}{4}-\dfrac{%
\mathbf{\Delta }\Phi }{2}.
\end{equation*}%
Now letting $h\rightarrow 0$ and using assumption 3 on $\Phi $\ we get 
\begin{equation*}
\left\| D_{i}u\right\| _{B_{\Phi }^{k+1}(\mathbb{R}^{\Lambda })}\leq C\left[
\left\| \hat{A}u\right\| _{B_{\Phi }^{k}(\mathbb{R}^{\Lambda })}+\left\|
u\right\| _{B_{\Phi }^{k}(\mathbb{R}^{\Lambda }))}+\left\| u\right\|
_{B_{\Phi }^{k+1}(\mathbb{R}^{\Lambda })}\right]
\end{equation*}%
it then follows that 
\begin{equation*}
D_{i}u\in B_{\Phi }^{k+1}(\mathbb{R}^{\Lambda }).
\end{equation*}%
It then only remains to prove that $Z_{\Phi }^{k+2}u\in L^{2}(\mathbb{R}%
^{\Lambda })$. To see this first observe that 
\begin{equation}
Z_{\Phi }^{2}u=\hat{A}u+\mathbf{\Delta }u+\dfrac{\mathbf{\Delta }\Phi }{2}u.
\end{equation}%
Here, the Laplacian is taken in the distributional sense. Multiplying by $%
Z_{\Phi }^{k}$ on both sides of this last equality, we obtain:%
\begin{equation}
Z_{\Phi }^{k+2}u=Z_{\Phi }^{k}\hat{A}u+Z_{\Phi }^{k}\mathbf{\Delta }%
u+Z_{\Phi }^{k}\dfrac{\mathbf{\Delta }\Phi }{2}u.
\end{equation}%
The first term of this equality is in $L^{2}(\mathbb{R}^{\Lambda })$ because 
$\hat{A}u\in B_{\Phi }^{k}(\mathbb{R}^{\Lambda }).$ That the second terms
also belongs to $L^{2}(\mathbb{R}^{\Lambda })$ follows from the fact that $%
D_{i}u\in B_{\Phi }^{k+1}(\mathbb{R}^{\Lambda })$. Finally to see that the
last term is an element of $L^{2}(\mathbb{R}^{\Lambda }),$ we use assumption
3 on $\Phi $ to get that 
\begin{eqnarray}
\dfrac{\mathbf{\Delta }\Phi }{2} &\leq &C(\frac{1}{4}+Z_{\Phi }^{2})^{1/2} \\
&\leq &C(\frac{1}{2}+Z_{\Phi }),
\end{eqnarray}%
and use the fact that $u\in B_{\Phi }^{k+1}(\mathbb{R}^{\Lambda }).$
\end{proof}

\begin{proposition}[$C^{\infty }-$regularity]
\textit{The weak solution }$u$ \textit{of }$W_{\Phi }^{(0)}u=q$\textit{\ is
an element of }$C^{\infty }(\mathbb{R}^{\Lambda }).$
\end{proposition}

The proof of this proposition use the general Sobolev inequalities theorem
given below.

\begin{theorem}[General Sobolev Inequality]
Let $U$ be a bounded open subset of $\mathbb{R}^{n},$ with a $C^{1}-$%
boundary. Assume $u\in W^{k,p}(U)$ where 
\begin{equation*}
W^{k,p}(U):=\left\{ u\in L_{loc}^{1}(\mathbb{R}^{n}):\partial ^{\alpha }u\in
L^{p}\left( \mathbb{R}^{n}\right) \mathbb{\ \ \ \forall }\left| \mathbb{%
\alpha }\right| \leq k\right\} .
\end{equation*}%
If 
\begin{equation*}
k>\frac{n}{p}
\end{equation*}%
then $u\in C^{k-\left[ \frac{n}{p}\right] -1,\gamma }(\bar{U}),$ where 
\begin{equation*}
\gamma =\left\{ 
\begin{tabular}{l}
$\left[ \frac{n}{p}\right] +1-\frac{n}{p},\ \ \ \ $if $\frac{n}{p}$ is not
an integer. \\ 
any positive number $<1,$ \ if $\frac{n}{p}$ is an integer.%
\end{tabular}%
\right.
\end{equation*}%
Here $C^{k,\alpha }(\bar{U})$ is the H\"{o}lder space consisting of all
functions $u\in C^{k}(\bar{U})$ such that 
\begin{equation*}
\left\| u\right\| _{C^{k,\alpha }(\bar{U})}:=\sum_{\left| \beta \right| \leq
k}\sup_{x\in U}\left| \partial ^{\beta }u(x)\right| +\sum_{\left| \beta
\right| =k}\sup_{\substack{ x,y\in U  \\ x\neq y}}\left| \frac{\partial
^{\beta }u(x)-\partial ^{\beta }u(y)}{\left| x-y\right| ^{\alpha }}\right|
<\infty
\end{equation*}
\end{theorem}

\begin{proof}
see [3]
\end{proof}

\begin{proof}[Proof of proposition 11]
Because $q\in C^{\infty }(\mathbb{R}^{\Lambda })$ , we have $u\in $ $%
B_{loc}^{k}(\mathbb{R}^{\Lambda })$ $\forall k,$ which implies $u\in
H^{k}(V)\left( =W^{k,2}(V)\right) $ $\forall k$ and $\forall V\subset
\subset \mathbb{R}^{\Lambda }.$ Now choose $k\in \mathbb{N}$ such that $%
k>\left\vert \Lambda \right\vert .$ Then the theorem $\ $above implies that $%
u\in C^{k,\gamma }(\bar{V})$ for some $0<\gamma <1$. Consequently, $u\in
C^{k}(V)$ for an arbitrary big enough $k$ and for any $V\subset \subset 
\mathbb{R}^{\Lambda }$
\end{proof}

Now that we have enough smoothness, we can make the following remark which
completes the proof of theorem 9.

\begin{remark}
A simple integration by parts argument shows that $u$ is in fact a strong
solution. It satisfies%
\begin{equation*}
W_{\Phi }^{(0)}u=q
\end{equation*}%
pointwise almost everywhere. Using the unitary transformation and taking
gradient on both sides of%
\begin{equation*}
A_{\Phi }^{(0)}v=g-\left\langle g\right\rangle _{_{L^{2}(\mu )}},
\end{equation*}%
we get 
\begin{equation*}
A_{\Phi }^{(1)}\mathbf{\nabla }v=\mathbf{\nabla }g.
\end{equation*}%
If $\mathbf{q}$\ is a smooth vector field satisfying 
\begin{equation}
\left| \partial ^{\alpha }\mathbf{q}\right| \leq C_{\alpha }(1+Z_{\Phi
})^{q_{\alpha }}\text{ \ \ for some\ }q_{\alpha }>0,
\end{equation}%
then one can show as above (this time using uniqueness result of the
Fredholm alternative) that the equation 
\begin{equation*}
A_{\Phi }^{(1)}\mathbf{v}=\mathbf{q}
\end{equation*}%
has a unique weak solution.\newline
$A_{\Phi }^{(1)}\mathbf{\nabla }v=\mathbf{\nabla }g$ would then imply that
two solutions of $\ $%
\begin{equation}
A_{\Phi }^{(0)}v=g-\left\langle g\right\rangle _{_{L^{2}(\mu )}}
\end{equation}%
must differ by a constant. Thus the problem 
\begin{equation*}
\left\{ 
\begin{tabular}{l}
$A_{\Phi }^{(0)}v=g-\left\langle g\right\rangle _{_{L^{2}(\mu )}}$ \\ 
$\left\langle v\right\rangle _{L^{2}(\mu )}=0$%
\end{tabular}%
\right.
\end{equation*}%
has a unique solution. This ends the proof of theorem 9.
\end{remark}

\section{The Kac-like Model}

In this section, we propose to illustrate the results above through the
study of a more specific family of classical unbounded spin model related to
Statistical Mechanics and is given by%
\begin{equation}
\Phi (x)=\Phi _{\Lambda }(x)=\frac{x^{2}}{2}+\Psi (x),\;\ \ \ \ \;x\in 
\mathbb{R}^{\Lambda }.
\end{equation}%
Here we have used the notation $x^{2}=x\cdot x.$\newline
The model that was originally suggested by M. Kac corresponds to when $\Psi $
is given by%
\begin{equation*}
\Psi (x)=-2\sum_{i,j\in \Lambda ,i\sim j}\ln \cosh \left[ \sqrt{\frac{\nu }{2%
}}\left( x_{i}+x_{j}\right) \right]
\end{equation*}%
where $\nu $ is a small positive constant.\newline
Other aspects of this family of potentials are studied in [8] in the one
dimensional case.

\begin{definition}
The lattice support, $S_{g}$ of a function $g$ on $\mathbb{R}^{\Lambda }$ is
defined to be the smallest subset $\Gamma $ of $\Lambda $ for which $g$ can
be written as function of $x_{l}$ alone with $l\in \Gamma .$ For instance,
if $g=x_{i},$ $S_{g}=\{i\}.$
\end{definition}

Under the assumptions 
\begin{equation}
\left| \partial ^{\alpha }\mathbf{\nabla }\Psi \right| \leq C_{\alpha },\;\
\ \ \ \ \forall \alpha \in \mathbb{N}^{\left| \Lambda \right| },\ 
\end{equation}%
\begin{equation}
\mathbf{Hess}\Phi \geq \delta >0,\;\;\;0<\delta <1,
\end{equation}%
One can check that $\Phi $ satisfies the assumptions 1-4 in section 3.

Let $g$ be a smooth function on $\mathbb{R}^{\Gamma }$ where $\Gamma $ is a
fixed subset. We shall use the notation 
\begin{equation*}
x_{\Sigma }=\left( x_{i}\right) _{i\in \Sigma }
\end{equation*}%
if $\Sigma $ is a proper subset of $\Lambda $ and shall also assume that $%
S_{g}=\Gamma .$ Now define the function $\tilde{g}$ on $\mathbb{R}^{\Lambda
} $ by 
\begin{equation*}
\tilde{g}(x)=g(x_{\Gamma }),\ \ \ \ \ \ \ x\in \mathbb{R}^{\Lambda }.
\end{equation*}%
If there is no ambiguity we shall identify $\tilde{g}$ with $g.$\newline
We propose to prove that if in addition to the assumptions above on $\Phi ,$
the functions $\Psi $ and $g$ are compactly supported and $g$ satisfies,%
\begin{equation*}
\left| \partial ^{\alpha }\mathbf{\nabla }g\right| \leq C_{\alpha },\;\ \ \
\ \ \forall \alpha \in \mathbb{N}^{\left| \Lambda \right| },
\end{equation*}

then the solution $v$ of the equation 
\begin{equation}
\left\{ 
\begin{tabular}{l}
$-\mathbf{\Delta }v+\mathbf{\nabla }\Phi \cdot \mathbf{\nabla }%
v=g-\left\langle g\right\rangle _{_{L^{2}(\mu )}}$ \\ 
$\left\langle v\right\rangle _{L^{2}(\mu )}=0$%
\end{tabular}%
\right. \text{in }\mathbb{R}^{\Lambda }
\end{equation}%
constructed in section 5 satisfies 
\begin{equation}
\partial ^{\alpha }\mathbf{\nabla }v(x)\rightarrow 0\;\text{as}\ \left|
x\right| \rightarrow \infty \ \ \ \ \forall \alpha \in \mathbb{N}^{\left|
\Lambda \right| }.
\end{equation}%
Recall that under a suitable change of variables, the equation 
\begin{equation}
A_{\Phi }^{(1)}v=\mathbf{\nabla }g
\end{equation}%
could\ be written as 
\begin{equation}
\left( \mathbf{-\Delta +}\frac{\left| \mathbf{\nabla }\Phi \right| ^{2}}{4}-%
\frac{\mathbf{\Delta }\Phi }{2}\right) \otimes \mathbf{u}+\mathbf{Hess}\Phi 
\mathbf{u}=\mathbf{q}
\end{equation}%
where 
\begin{equation}
\mathbf{u}=e^{-\Phi /2}\mathbf{\nabla }v\ \ \ \ \text{\ and}\ \ \ \mathbf{\ q%
}=e^{-\Phi /2}\mathbf{\nabla }g
\end{equation}%
Let $B_{1}=B_{R_{1}}(0)\subset \mathbb{R}^{\Lambda }$denote a large balls
centered at zero with radius $R_{1}$ and containing \ the support of $\Psi $
in $\mathbb{R}^{\Lambda }.$ We also consider a ball $B_{2}=B_{R_{2}}(0)%
\subset \mathbb{R}^{\Gamma }$ of radius $R_{2}>R_{1}$ containing the support
of $g$ in $\mathbb{R}^{\Gamma }.$ The support of $\tilde{g}$ in $\mathbb{R}%
^{\Lambda }$ is then contained in the cylinder%
\begin{equation*}
B=B_{2}\times \mathbb{R}^{\Lambda \backslash \Gamma }.
\end{equation*}%
In $B^{c}=\mathbb{R}^{\Lambda }\backslash B$ we have%
\begin{equation}
\left\{ 
\begin{tabular}{l}
$\left( \mathbf{-\Delta +}\dfrac{x^{2}}{4}-\dfrac{m}{2}+\mathbf{I}\right) 
\mathbf{u}=0$ \ \ in $B^{c}$ \\ 
$\mathbf{u}=\mathbf{\varphi }$ \ \ \ \ \ \ \ \ \ \ \ \ \ \ \ \ \ \ \ \ \ on $%
\partial B\;($in the trace sense$).$%
\end{tabular}%
\right.
\end{equation}%
Here $\mathbf{\varphi }$ is a $C^{\infty }-$vector field on $\partial B$ and 
$m=\left| \Lambda \right| .$

Since the operator 
\begin{equation}
\mathbf{-\Delta +}\dfrac{x^{2}}{4}-\dfrac{m}{2}+\mathbf{I}
\end{equation}%
acts diagonally on\textbf{\ }$\mathbf{u},$ we can work component by
component and the situation is reduced to the scalar case 
\begin{equation}
\left\{ 
\begin{tabular}{l}
$\left( \mathbf{-\Delta +}\dfrac{x^{2}}{4}-\dfrac{m}{2}+1\right) u=0$ \ \ in 
$B^{c}$ \\ 
$u=\varphi $ \ \ \ \ \ \ \ \ \ \ \ \ \ \ \ \ \ \ \ \ \ on $\partial B\;($in
the trace sense$).$%
\end{tabular}%
\right. .
\end{equation}%
\newline
Having reduced the problem to a Dirichlet type for the Schrodinger operator%
\begin{equation}
\mathbf{-\Delta +}\dfrac{x^{2}}{4}-\dfrac{m}{2}+1,
\end{equation}%
we shall need some results on the decay of eigenfunctions of the
corresponding Schrodinger operator. We need the following lemma:

\begin{lemma}
\textit{The fundamental solution }$\mathcal{E}\in \mathcal{S}^{\prime }(%
\mathbb{R}^{\Lambda })$ \textit{of the operator }$-\mathbf{\Delta }%
+k^{2}\;(k>0)$\textit{\ exists and is unique. It is spherically symmetric,
is an element of}\ $C^{\infty }(\mathbb{R}^{\Lambda }\backslash \{0\})$ a%
\textit{nd has the following asymptotics as }$\left\vert x\right\vert
\rightarrow \infty :$\textit{\ }%
\begin{equation}
\mathcal{E}(x)=C\left\vert x\right\vert ^{\left( \frac{m-1}{2}\right)
}e^{-k\left\vert x\right\vert }(1+o(1))
\end{equation}%
\newline
In the Lemma, $\mathcal{S}^{\prime }(\mathbb{R}^{\Lambda })$ denotes the
space of tempered distributions on $\mathbb{R}^{\Lambda }$
\end{lemma}

\begin{proof}
Consider the equation 
\begin{equation}
\left( -\mathbf{\Delta }+k^{2}\right) \mathcal{E}(x)=\delta _{o}(x).
\end{equation}%
Taking Fourier transform, we get 
\begin{equation}
\widehat{\left( -\mathbf{\Delta }+k^{2}\right) \mathcal{E}(x)}=\widehat{%
\delta _{o}(x)}.
\end{equation}%
equivalently%
\begin{equation}
\left( x^{2}+k^{2}\right) \widehat{\mathcal{E}(x)}=\left( 2\pi \right)
^{-m/2}
\end{equation}%
which implies 
\begin{equation}
\widehat{\mathcal{E}(x)}=\frac{\left( 2\pi \right) ^{-m/2}}{x^{2}+k^{2}}.
\end{equation}%
The uniqueness and spherical symmetry follow since$\ $%
\begin{equation}
\mathcal{E}(x)=\left( 2\pi \right) ^{-m/2}\widehat{\widehat{\mathcal{E}(x)}}.
\end{equation}%
Furthermore, if $x\neq 0,$ the smoothness of $\mathcal{E}(x)$ follows from
the regularity theory of the elliptic equation as discussed above in section
5.%
\begin{equation}
\left( -\mathbf{\Delta }+k^{2}\right) \mathcal{E}(x)=0\;\ \text{in\ \ }%
\mathbb{R}^{\Lambda }\backslash \{0\}
\end{equation}%
for $x\neq 0$ set $\mathcal{E}(x)=f(r)$ where $f\in C^{\infty }(\mathbb{R}%
^{+})$ and $r=\left| x\right| .$ $(91)$ becomes 
\begin{equation}
-f^{\;\prime \prime }(r)-\frac{m-1}{r}f^{\;\prime }(r)+k^{2}f(r)=0.
\end{equation}%
Set $f(r)=a(r)g(r).$ Plugging this in $(92)$ and setting the coefficient of $%
g^{\prime }(r)$ equal zero gives 
\begin{equation}
2a^{\prime }+\frac{m-1}{r}a=0.
\end{equation}%
Take%
\begin{equation*}
a(r)=r^{-\tfrac{m-1}{2}}.
\end{equation*}%
Then%
\begin{equation*}
f(r)=r^{-\tfrac{m-1}{2}}g(r)
\end{equation*}%
and $(92)$ takes the form%
\begin{equation}
g^{\prime \prime }(r)-k^{2}(1+O(\frac{1}{r^{2}}))g(r)=0
\end{equation}%
Now using classical results on the asymptotics of the solutions of the
Schrodinger operator (see [65]), we discover that 
\begin{equation}
g_{\pm }(r)=Ce^{\pm kr}(1+o(1)).
\end{equation}%
Hence the asymptotics of the solutions of $(92)$ are%
\begin{equation}
f_{\pm }(r)=Cr^{-\tfrac{m-1}{2}}e^{\pm kr}(1+o(1)).
\end{equation}%
Since $\mathcal{E}(x)=f(\left| x\right| )\in \mathcal{S}^{\prime }(\mathbb{R}%
^{\Lambda }),$ we conclude that $f=f_{-}$ and the result\ follows.
\end{proof}

\begin{theorem}
\textit{Let }$\Omega $\textit{\ be any exterior domain in }$\mathbb{R}%
^{\Lambda }$ \textit{containing a neighborhood of infinity with smooth
internal boundary.} \textit{Let the potential }$v(x)\in C^{\infty }(\Omega )$
\textit{and} \textit{satisfy }%
\begin{equation}
\lim\limits_{\left| x\right| \rightarrow \infty }\inf v(x)\geq E
\end{equation}%
\textit{and let }$\varphi $\textit{\ be a smooth solution of the problem }%
\begin{equation}
\left\{ 
\begin{tabular}{ll}
$\left( -\mathbf{\Delta }+v(x)\right) \varphi =\lambda \varphi $ & \ \ in $%
\Omega $ \\ 
$\rho =\psi $ & \ \ on $\partial \Omega $%
\end{tabular}%
\right.
\end{equation}%
\textit{where }$\lambda <E$\textit{\ and }$\varphi $ is a smooth function on 
$\partial \Omega .$\textit{\ Then the following estimate holds: }%
\begin{equation}
\left| \varphi (x)\right| \leq C_{\varepsilon }e^{-\sqrt{\left( a-\lambda
-\varepsilon \right) /2}\left| x\right| }
\end{equation}%
\textit{for any }$\varepsilon >0.$
\end{theorem}

The proof of this theorem uses the following lemma

\begin{lemma}[A Maximum principle]
\textit{Let }$k>0,$\textit{\ }$\Sigma $\textit{\ an open subset of} $\mathbb{%
R}^{\Lambda },$ \textit{and }$u\in C^{2}(\Sigma )$\textit{\ a function such
that }%
\begin{equation}
\left( -\mathbf{\Delta }+k^{2}\right) u=f\leq 0\;\;\;\;\;\text{in}\;\Sigma .
\end{equation}%
\textit{Then }$u$\textit{\ cannot have a positive maximum in} $\Sigma $.
\end{lemma}

\begin{proof}
If $x_{o}\in \Sigma $ is a maximum point and $u(x_{o})>0,$ then 
\begin{equation}
\mathbf{\Delta }u(x_{o})\leq 0;
\end{equation}%
this contradicts $(100).$
\end{proof}

\begin{proof}[Proof of Theorem 7]
Let $\varphi $ be a real solution of the equation 
\begin{equation}
H\varphi =\lambda \varphi \text{ \ \ \ in }\Omega .
\end{equation}%
where 
\begin{equation*}
H=-\mathbf{\Delta }+v(x).
\end{equation*}%
We obviously have\ 
\begin{equation}
\mathbf{\Delta }\left( \mathbf{\varphi }^{2}\right) =2\mathbf{\Delta \varphi
\cdot \varphi +}2\left| \mathbf{\nabla }\varphi \right| ^{2}
\end{equation}%
$H\varphi =\lambda \varphi $ gives $-\mathbf{\Delta }\varphi =\left( \lambda
-v(x)\right) \varphi $ which implies 
\begin{equation}
\mathbf{\Delta }\left( \mathbf{\varphi }^{2}\right) =2\left( \lambda
-v(x)\right) \varphi ^{2}-2\left| \mathbf{\nabla }\varphi \right| ^{2}
\end{equation}%
adding $2(b-\lambda )\varphi ^{2}$ on both sides of this equality, we obtain 
\begin{equation}
\left[ -\mathbf{\Delta +}2(b-\lambda )\right] \varphi ^{2}=-2\left(
v(x)-b\right) \varphi ^{2}-2\left| \mathbf{\nabla }\varphi \right| ^{2}.
\end{equation}%
Choosing $\lambda <b<E$ the right hand side of $(105)$ is non-positive for $%
\left| x\right| $ large enough.  Now set 
\begin{equation}
u(x)=\varphi ^{2}(x)-M\mathcal{E}(x)
\end{equation}%
where $\mathcal{E}(x)$ is the fundamental solution of the operator $-\mathbf{%
\Delta }+k^{2}$ with%
\begin{equation}
k=\sqrt{2(b-\lambda )}.
\end{equation}%
Choose $R$ so large that $\mathcal{E}(x)>0$ and $v(x)>b$ for $\left|
x\right| >R.$ Now choose $M$ so large that $u(x)<0$ on $\left\{ x\in 
\overline{\Omega }:\left| x\right| =R\right\} .$ We shall prove that 
\begin{equation}
u(x)\leq 0
\end{equation}%
on $\left\{ x\in \overline{\Omega }:\left| x\right| =R\right\} $\ from which
the theorem will follow.  Substracting from $(105)$ the equation 
\begin{equation}
\left[ -\mathbf{\Delta +}2(b-\lambda )\right] M\mathcal{E}(x)=0,
\end{equation}%
we find that $(100)$ is satisfied for $u(x)$ with%
\begin{equation}
f=-2\left( v(x)-b\right) \varphi ^{2}-2\left| \mathbf{\nabla }\varphi
\right| ^{2},\;\;\;\;\;\;\;\text{for }\left| x\right| \geq R.
\end{equation}%
We then apply the maximum principle in each connected component of the subset%
\begin{equation}
\Omega _{R,\rho }=\left\{ x\in \overline{\Omega }:R\leq \left| x\right| \leq
\rho \right\}
\end{equation}%
to the function 
\begin{equation}
u^{\varepsilon }(x)=\int u(x-y)\eta _{\varepsilon }(y)dy
\end{equation}%
where $\eta _{\varepsilon }(x)=\varepsilon ^{-m}\eta (\dfrac{x}{\varepsilon }%
)$ and $\eta (x)$ is the mollifier. Recall that $\eta (x)$ is given by 
\begin{equation*}
\eta (x)=\left\{ 
\begin{tabular}{l}
$e^{-\left( \tfrac{1}{1-\left| x\right| ^{2}}\right) }$ \ \ \ if $\left|
x\right| \leq 1$ \\ 
$0$ \ \ \ \ \ \ \ \ \ \ \ \ \ \ \ otherwise.%
\end{tabular}%
\right.
\end{equation*}%
We indeed have 
\begin{equation}
\left( -\mathbf{\Delta }+k^{2}\right) u^{\varepsilon }=f^{\varepsilon }=\int
f(x-y)\eta _{\varepsilon }(y)dy\leq 0
\end{equation}%
$u\in L^{1}(\mathbb{R}^{\Lambda })$ implies that $u^{\varepsilon
}(x)\rightarrow 0$ as $\left| x\right| \rightarrow \infty .$ Set 
\begin{equation}
M_{\rho }(\varepsilon )=\max\limits_{\left\{ x\in \overline{\Omega }:\left|
x\right| =\rho \right\} }\left| u^{\varepsilon }(x)\right|
\end{equation}%
since $u(x)<0$ for $x\in \left\{ x\in \overline{\Omega }:\left| x\right|
=R\right\} ,$ using the fact that $u^{\varepsilon }(x)\rightrightarrows u(x)$
as $\varepsilon \rightarrow 0$ on $\left\{ x\in \overline{\Omega }:\left|
x\right| =R\right\} ,$ we conclude that $u^{\varepsilon }(x)<0$ on $\left\{
x\in \overline{\Omega }:\left| x\right| =R\right\} $ for small $\varepsilon $%
It then follows from lemma 17 that 
\begin{equation}
u^{\varepsilon }(x)\leq M_{\rho }(\varepsilon )\text{ \ \ for }x\in \Omega
_{R,\rho }.
\end{equation}%
Letting $\rho \rightarrow \infty ,$ we get 
\begin{equation}
u^{\varepsilon }(x)\leq 0\;\;\text{for }x\in \overline{\Omega }\text{ and }%
\left| x\right| \geq R.
\end{equation}%
Now since 
\begin{equation}
u^{\varepsilon }(x)\rightrightarrows u(x)\;\text{as\ }\varepsilon
\rightarrow 0
\end{equation}%
in every relatively compact subset of $\left\{ x\in \overline{\Omega }%
:\left| x\right| \geq R\right\} $, it follows that%
\begin{equation}
u(x)\leq 0\;\;\text{for }x\in \left\{ x\in \overline{\Omega }:\left|
x\right| \geq R\right\} .
\end{equation}
\end{proof}

\begin{corollary}
\textit{If }$v(x)\rightarrow \infty $\textit{\ as }$\left\vert x\right\vert
\rightarrow \infty ,$\textit{\ then for any eigenfunction }$\varphi $ 
\textit{of the boundary value problem in theorem 7 satisfies, the following
estimate}%
\begin{equation}
\left\vert \varphi (x)\right\vert \leq C_{a}e^{-a\left\vert x\right\vert }
\end{equation}%
\textit{where }$a>0$\textit{\ is arbitrary and }$C_{a}>0.$
\end{corollary}

\begin{theorem}[{Helffer-Sj\"{o}strand [8]}]
\textit{The }$L^{2}-$\textit{solution }$u$\textit{\ of }%
\begin{equation}
\left\{ 
\begin{tabular}{l}
$\left( \mathbf{-\Delta +}\dfrac{x^{2}}{4}-\dfrac{m}{2}+1\right) u=0$\textit{%
\ \ \ in }$B^{c}$ \\ 
$u=\varphi $\textit{\ \ \ \ \ \ \ \ \ \ \ \ \ \ \ \ \ \ \ \ \ \ on }$%
\partial B\;($\textit{in the trace sense}$).$%
\end{tabular}%
\right. (E)
\end{equation}%
\textit{satisfies}%
\begin{equation}
u(x)=e^{-\tfrac{x^{2}}{4}}\left| x\right| ^{-1/2}h(x)
\end{equation}%
\textit{where} 
\begin{equation}
\partial ^{\beta }h(x)=O(\left| x\right| ^{-\left| \beta \right|
})\;\;\;\forall \beta \in \mathbb{N}^{m}.
\end{equation}%
Using the change of variable $v=e^{\Phi /2}u$ and applying this theorem to
each component of $u,$ we obtain
\end{theorem}

\begin{corollary}
\textit{The }$L^{2}-$\textit{solution }$v$\textit{\ of the system }%
\begin{equation}
\left( -\mathbf{\Delta }+\mathbf{\nabla }\Phi \cdot \mathbf{\nabla }\right) 
\mathbf{v}+\mathbf{Hess}\Phi \mathbf{v}=\mathbf{\nabla }g\;\;\;\text{in}\;%
\mathbb{R}^{\Lambda }
\end{equation}%
\textit{satisfies} 
\begin{equation}
\lim\limits_{\left| x\right| \rightarrow \infty }\partial ^{\alpha }\mathbf{v%
}(x)=0\;\;\;\;\forall \alpha \in \mathbb{N}^{m}.
\end{equation}
\end{corollary}

\begin{proof}[{Proof of theorem 8 (Sj\"{o}strand [14])}]
Denote by 
\begin{equation}
K:C^{\infty }(\partial B)\rightarrow C^{\infty }\left( B^{c}\right)
\end{equation}%
the operator that assigns each boundary value the corresponding solution.
Since by theorem 16%
\begin{equation}
\lim\limits_{\left| x\right| \rightarrow \infty }u(x)=0,
\end{equation}%
the maximum principle implies that $K$ is monotone increasing. Indeed, $%
Kg\geq 0$ whenever $g\geq 0.$ This implies that the operator $K$ is
increasing and that $Kg\leq \sup g,$ if $\sup g\geq 0,$ $Kg\geq \inf g$ if $%
\ \inf g\leq 0.$ Let 
\begin{equation}
u_{o}=K1\;\left( \geq 0\right)
\end{equation}%
which is a radial function i.e. 
\begin{equation}
u_{o}(x)=u_{o}(\left| x\right| );
\end{equation}%
with 
\begin{equation}
\left[ -\partial _{r}^{2}-\left( \dfrac{m-1}{r}\right) \partial _{r}+\dfrac{%
r^{2}}{4}-\dfrac{m}{2}+1\right] u_{o}(r)=0,\;\;\;\;\;\;u_{o}(R)=1.
\end{equation}%
We perform the Liouville's transformation 
\begin{equation}
u_{o}=r^{-(m-1)/2}f(r)
\end{equation}%
to get rid of the term involving $\partial _{r}.$ We finally get 
\begin{equation}
\left[ -\partial _{r}^{2}+\dfrac{r^{2}}{4}-\dfrac{\left( m-1\right) \left(
m-3\right) }{4r^{2}}+1-\dfrac{m}{2}\right] f(r)=0,\;\;\;\;\;%
\;f(R)=R^{(m-1)/2}.
\end{equation}%
which we write in the form 
\begin{equation}
\left[ -\partial _{r}^{2}+V(r)\right] f(r)=\dfrac{m}{2}f(r),\;\;\;\;\;%
\;f(R)=R^{(m-1)/2}.
\end{equation}%
where 
\begin{equation}
V(r)=\dfrac{r^{2}}{4}-\dfrac{\left( m-1\right) \left( m-3\right) }{4r^{2}}%
+1\rightarrow \infty \;\;\text{as}\;\;r\rightarrow \infty .
\end{equation}%
Since 
\begin{equation}
\int_{r_{o}}^{\infty }\frac{\left| V^{\prime }(r)\right| ^{2}}{\left|
V(r)\right| ^{5/2}}dr<\infty \text{ \ and }\int_{r_{o}}^{\infty }\frac{%
\left| V^{\prime \prime }(r)\right| ^{2}}{\left| V(r)\right| ^{3/2}}%
dr<\infty \text{ \ for some large }r_{o}.
\end{equation}%
Classical results on Schrodinger operators ( see [65] ) allow us to get the
asymptotics of $f(r)$ as following:%
\begin{equation}
f_{\pm }(r)=Cr^{-1/2}e^{\pm \tfrac{r^{2}}{4}}(1+o(1)).
\end{equation}%
Now since $u_{o}\rightarrow 0$ as $r\rightarrow \infty ,$ we conclude that 
\begin{equation}
f(r)=f_{-}(r)=Cr^{-1/2}e^{-\tfrac{r^{2}}{4}}(1+o(1)).
\end{equation}%
Hence 
\begin{equation}
u_{o}(r)=Cr^{-\tfrac{m}{2}}e^{-\tfrac{r^{2}}{4}}(1+o(1))>0.
\end{equation}%
Next, we write 
\begin{equation}
u(x)=j(x)u_{o}(r)
\end{equation}%
Let $g\in C^{\infty }(\partial B)$ be strictly positive everywhere and let 
\begin{equation}
u=Kg.
\end{equation}%
Denote by $g_{\min }=\inf g$ and $g_{\max }=\sup g.$ We obviously have 
\begin{equation}
g_{\min }u_{o}\leq u\leq g_{\max }u_{o}.
\end{equation}%
Hence, 
\begin{equation}
j(x)=\dfrac{u(x)}{u_{o}(x)}
\end{equation}%
is bounded. Next, we perform a change in polar coordinates $(r,\theta )$ by
setting $x=r\theta .$ Under this change of coordinates, the operator 
\begin{equation}
\mathbf{-\Delta +}\dfrac{x^{2}}{4}-\dfrac{m}{2}+1
\end{equation}%
becomes 
\begin{equation}
-\partial _{r}^{2}-\left( \dfrac{m-1}{r}\right) \partial _{r}+\dfrac{r^{2}}{4%
}-\dfrac{m}{2}+1-r^{-2}\mathbf{\Delta }_{\theta }
\end{equation}%
where $\mathbf{\Delta }_{\theta }$ is the Laplace-Beltrami operator on $%
S^{m-1}.$  Since the operator $\mathbf{-\Delta +}\dfrac{x^{2}}{4}-\dfrac{m}{2%
}+1$ is rotationally invariant and $\partial _{\theta }^{\alpha }u$ takes
continuously the value $\partial _{\theta }^{\alpha }g$ on $\partial B,$
using the fact that each $\partial _{\theta }^{\alpha }u$ arises as
infinitesimal rotation, we conclude that for every $\alpha ,$ $\partial
_{\theta }^{\alpha }u$ is a solution of the boundary value problem (E)
(under the change of coordinates) with 
\begin{equation}
\partial _{\theta }^{\alpha }u=\partial _{\theta }^{\alpha }g\;\;\;\text{on}%
\;\;\partial B.
\end{equation}%
Therefore, 
\begin{equation}
\partial _{\theta }^{\alpha }u=O(1)e^{-\tfrac{r^{2}}{4}},\;\;\;\;\;\forall
\alpha \in \mathbb{N}^{m},
\end{equation}%
which implies 
\begin{equation}
\partial _{\theta }^{\alpha }j=O(1),\;\;\;\;\forall \alpha \in \mathbb{N}%
^{m}.
\end{equation}%
Now we need to control some radial derivative of $j.$ In polar coordinates,
we have 
\begin{equation}
\left[ -\partial _{r}^{2}-\left( \dfrac{m-1}{r}\right) \partial _{r}+\dfrac{%
r^{2}}{4}-\dfrac{m}{2}+1-r^{-2}\mathbf{\Delta }_{\theta }\right] u_{o}(r)=0.
\end{equation}%
Write%
\begin{equation}
\left[ -\partial _{r}^{2}-\left( \dfrac{m-1}{r}\right) \partial _{r}+\dfrac{%
r^{2}}{4}-\dfrac{m}{2}+1-r^{-2}\mathbf{\Delta }_{\theta }\right] j(r,\theta
)u_{o}(r)=0.
\end{equation}%
Using $(129)$ and the product rule of differentiation, $(148)$ becomes \ 
\begin{equation}
\left[ \partial _{r}^{2}+\left[ 2\dfrac{\partial _{r}u_{o}}{u_{o}}+\left( 
\dfrac{m-1}{r}\right) \right] \partial _{r}\right] j=-r^{-2}\mathbf{\Delta }%
_{\theta }j.
\end{equation}%
Here 
\begin{equation}
\partial _{\theta }^{\alpha }\left( r^{-2}\mathbf{\Delta }_{\theta }j\right)
=O(r^{-2}),\;\;\;\;\forall \alpha \in \mathbb{N}^{m},
\end{equation}%
and 
\begin{equation}
\dfrac{\partial _{r}u_{o}}{u_{o}}=-\dfrac{r}{2}+O(\dfrac{1}{r}).
\end{equation}%
Thus, $\left( 149\right) $ can be written as 
\begin{equation}
\left[ \partial _{r}^{2}+\left[ -r+O(\dfrac{1}{r})\right] \partial _{r}%
\right] j=O(r^{-2}).
\end{equation}%
Let 
\begin{equation}
\varphi (r)=r+O(\dfrac{1}{r}).
\end{equation}%
We have 
\begin{equation}
\left[ \partial _{r}-f(r)\right] \partial _{r}j=O(r^{-2}).
\end{equation}%
Let 
\begin{equation}
F(r)=\int_{1}^{r}f(t)dt\sim r^{2}.
\end{equation}%
Solving $\left( 154\right) ,$ we get 
\begin{equation}
\partial _{r}j=-\int_{r}^{\infty }e^{F(r)-F(s)}\left[ O(s^{-2})\right]
ds+Ce^{F(r)}.
\end{equation}%
Since%
\begin{equation}
F(r)-F(s)\sim r^{2}-s^{2}\leq 2r(r-s)\;\;\text{for}\;\;s\geq r,
\end{equation}%
$\partial _{r}j$ cannot tend to $\pm \infty $ when $r\rightarrow \infty ,$
we conclude that $C=0$ and 
\begin{equation}
\partial _{r}j=-\int_{r}^{\infty }e^{F(r)-F(s)}\left[ O(s^{-2})\right]
ds=O(r^{-3}).
\end{equation}%
More generally, since $\partial _{\theta }^{\alpha }j$ is a solution of $%
\left( 157\right) $ with right hand side%
\begin{equation}
-r^{-2}\partial _{\theta }^{\alpha }\left( \mathbf{\Delta }_{\theta
}j\right) =O(r^{-2}),
\end{equation}%
using the same argument as above with $j$ replaced by $\partial _{\theta
}^{\alpha }j,$ we have 
\begin{equation}
\partial _{r}\partial _{\theta }^{\alpha }j=O(r^{-3}).
\end{equation}%
Now differentiating 
\begin{equation}
\left[ \partial _{r}-f(r)\right] \partial _{r}\partial _{\theta }^{\alpha
}j=O(r^{-2})
\end{equation}%
with respect to $r,$ we get 
\begin{equation}
\left[ \partial _{r}-f(r)\right] \partial _{r}^{2}\partial _{\theta
}^{\alpha }j=O(r^{-3}),
\end{equation}%
using again the same argument as before, we get 
\begin{equation}
\partial _{r}^{2}\partial _{\theta }^{\alpha }j=O(r^{-4})
\end{equation}%
continuing this way, we finally get 
\begin{equation}
\partial _{r}^{k}\partial _{\theta }^{\alpha }j=O(r^{-2-k})\;\;\;\;k=1,2,...
\end{equation}%
Going back to $x-$coordinates, we get 
\begin{equation}
\partial ^{\alpha }j(x)=O(\left| x\right| ^{-\left| \alpha \right|
}),\;\;\;\;\forall \alpha \in \mathbb{N}^{m},\alpha \neq 0.
\end{equation}
\end{proof}

\section{Weighted Estimates for the Decay of Correlation}

In this section, we propose to get estimates suitable for obtaining the
decay of the correlation functions. We shall first analyze the case where $%
\Psi $ and the source term $g$ are compactly supported

\subsection{The compactly supported case.}

We shall assume that $\Phi $ is given by 
\begin{equation}
\Phi (x)=\Phi _{\Lambda }(x)=\frac{x^{2}}{2}+\Psi (x),\;\ \ \ \ \;x\in 
\mathbb{R}^{\Lambda }.
\end{equation}%
where%
\begin{equation}
\left| \partial ^{\alpha }\mathbf{\nabla }\Psi \right| \leq C_{\alpha },\;\
\ \ \ \ \forall \alpha \in \mathbb{N}^{\left| \Lambda \right| }.
\end{equation}%
Again $g$ will denote a smooth function on $\mathbb{R}^{\Gamma }$ with
lattice support $S_{g}=\Gamma .$ We shall identify $g$ with $\tilde{g}$
defined on $\mathbb{R}^{\Lambda }$ and shall assume that%
\begin{equation}
\left| \partial ^{\alpha }\mathbf{\nabla }g\right| \leq C_{\alpha
}\;\;\;\;\;\;\;\;\forall \alpha \in \mathbb{N}^{\left| \Gamma \right| }.
\end{equation}%
In addition, we shall momentarily assume that $\Psi $ is compactly supported
in $\mathbb{R}^{\Lambda }$ and $g$ is compactly supported in $\mathbb{R}%
^{\Gamma }$ but these assumptions will be relaxed later on. Let $M$ be the
diagonal matrix%
\begin{equation*}
M=\left( \delta _{ij}\rho (i)\right) _{i,j\in \Lambda }
\end{equation*}%
where $\rho $ is a weight function on $\Lambda $ satisfying \ 
\begin{equation}
e^{-\lambda }\leq \frac{\rho \left( i\right) }{\rho (j)}\leq e^{\lambda }%
\text{, \ \ if }i\sim j\text{\ \ for some }\lambda >0.
\end{equation}%
Assume also that there exists $\delta _{o}\in (0,1)$ such that%
\begin{equation}
M^{-1}\mathbf{Hess}\Phi (x)M\geq \delta _{o}
\end{equation}%
for every $M$ as above.\newline
Let 
\begin{equation}
\rho (i)=e^{\kappa d(i,S_{g})}
\end{equation}%
where $\kappa $ is a positive. Define 
\begin{equation*}
\left| x\right| _{2,\rho }:=\left( \sum_{i\in \Lambda }\rho
(i)^{2}x_{i}^{2}\right) ^{1\backslash 2}.
\end{equation*}%
Let $f$ be the solution of the equation%
\begin{equation*}
\left\{ 
\begin{tabular}{l}
$-\mathbf{\Delta }f+\mathbf{\nabla }\Phi \cdot \mathbf{\nabla }%
f=g-\left\langle g\right\rangle $ \\ 
$\left\langle f\right\rangle _{L^{2}(\mu )}=0.$%
\end{tabular}%
\right.
\end{equation*}%
Recall that $\mathbf{\nabla }f$ is a solution of the system%
\begin{equation}
\left( -\mathbf{\Delta +\nabla }\Phi \cdot \mathbf{\nabla }\right) \mathbf{%
\nabla }f+\mathbf{Hess}\Phi \mathbf{\nabla }f=\mathbf{\nabla }g\;\;\;\;\text{%
in }\;\mathbb{R}^{\Lambda }.
\end{equation}%
Let $t_{1}=\left( t_{i}\right) _{i}\in \mathbb{R}^{\Lambda }$%
\begin{eqnarray}
\left\langle \mathbf{\nabla }\left( \mathbf{\nabla }\Phi \cdot \mathbf{%
\nabla }f\right) ,t_{1}\right\rangle &=&\sum\limits_{i,k\in \Lambda }\left(
f_{x_{i}}\Phi _{x_{i}x_{k}}t_{k}+\Phi _{x_{i}}f_{x_{i}x_{k}}t_{k}\right) \\
&=&\left\langle \mathbf{\nabla }f,\mathbf{Hess}\Phi t_{1}\right\rangle +%
\mathbf{\nabla }\Phi \cdot \mathbf{\nabla }\left\langle \mathbf{\nabla }%
f,t_{1}\right\rangle .
\end{eqnarray}%
On the other hand, \newline
\begin{equation*}
\left\langle \mathbf{\nabla }\left( \mathbf{\Delta }f\right)
,t_{1}\right\rangle =\mathbf{\Delta }\left\langle \mathbf{\nabla }%
f,t_{1}\right\rangle .
\end{equation*}%
We therefore have%
\begin{equation}
\left\langle \mathbf{\nabla }g,t_{1}\right\rangle =\left( \mathbf{\nabla }%
\Phi \cdot \mathbf{\nabla }-\mathbf{\Delta }\right) \left\langle \mathbf{%
\nabla }f,t_{1}\right\rangle +\left\langle \mathbf{\nabla }f,\mathbf{Hess}%
\Phi t_{1}\right\rangle .
\end{equation}%
Because $\mathbf{\nabla }f(x)\rightarrow 0$ as $\left| x\right| \rightarrow
\infty ,$ we consider a point $x_{o}$ at which \ 
\begin{equation*}
\left| \mathbf{\nabla }f(x)\right| _{2,\rho }=\left( \sum_{i\in \Lambda
}\rho (i)^{2}f_{x_{i}}^{2}(x)\right) ^{1\backslash 2}
\end{equation*}%
is maximal. If $M$ is the diagonal matrix 
\begin{equation*}
M=\left( \delta _{ij}\rho (i)\right)
\end{equation*}%
we have%
\begin{equation}
\left\langle \mathbf{\nabla }g,Mt_{1}\right\rangle =\left( \mathbf{\nabla }%
\Phi \cdot \mathbf{\nabla }-\mathbf{\Delta }\right) \left\langle \mathbf{%
\nabla }f,Mt_{1}\right\rangle +\left\langle \mathbf{\nabla }f,\mathbf{Hess}%
\Phi Mt_{1}\right\rangle .
\end{equation}%
Now choose 
\begin{equation*}
t_{1}=\left( \rho (i)f_{x_{i}}(x_{o})\right) _{i\in \Lambda }.
\end{equation*}%
We need the following lemma.

\begin{lemma}
\textit{Under the assumptions and notations above, the function }%
\begin{equation*}
x\longmapsto \left\langle \mathbf{\nabla }f(x),Mt_{1}\right\rangle
\end{equation*}%
\textit{achieves its maximum value at }$x_{o}.$
\end{lemma}

\begin{proof}
Let%
\begin{equation}
\zeta (x)=\left\langle \mathbf{\nabla }f(x),Mt_{1}\right\rangle
\end{equation}%
and 
\begin{equation}
\pi (x)=\left| \mathbf{\nabla }f(x)\right| _{2,\rho }^{2}.
\end{equation}%
Again by the maximum principle, the function $\zeta (x)$ achieves its
maximum at some $\bar{x}_{o}\in \mathbb{R}^{\Lambda }.$ It is easy to see
that $x_{o}$ is a critical point for $\zeta (x).$Moreover, for any $a\in 
\mathbb{R}^{\Lambda },$ we have%
\begin{eqnarray}
&&\left\langle a,\mathbf{Hess}\pi (x_{o})a\right\rangle \\
&=&2\left\langle a,\mathbf{Hess}\zeta (x_{o})a\right\rangle
+2\sum_{j,k}\left( \sum_{i}f_{x_{i}x_{j}}(x_{o})f_{x_{i}x_{k}}(x_{o})\rho
(i)^{2}\right) a_{j}a_{k} \\
&=&2\left\langle a,\mathbf{Hess}\zeta (x_{o})a\right\rangle +2\sum_{i}\rho
(i)^{2}\left( \sum_{j}f_{x_{i}x_{j}}(x_{o})\right) ^{2}
\end{eqnarray}%
Because $\left\langle a,\mathbf{Hess}\pi (x_{o})a\right\rangle <0,$ we must
have $\left\langle a,\mathbf{Hess}\zeta (x_{o})a\right\rangle <0$ for any $%
a\in \mathbb{R}^{\Lambda }.$ Thus, $x_{o}$ is a local maximum for $\zeta
(x). $Moreover, on one hand, we have 
\begin{equation}
\zeta (\bar{x}_{o})\geq \zeta (x_{o})=\pi (x_{o}).
\end{equation}%
One the other hand, Cauchy-Schwartz gives%
\begin{eqnarray}
\zeta (\bar{x}_{o}) &\leq &\left[ \pi (\bar{x}_{o})\right] ^{1/2}\left[ \pi
(x_{o})\right] ^{1/2} \\
&\leq &\pi (x_{o}).
\end{eqnarray}%
These last two above inequalities imply 
\begin{equation}
\zeta (\bar{x}_{o})=\zeta (x_{o})
\end{equation}%
and the result follows.
\end{proof}

Now using lemma 21 above, we have%
\begin{equation*}
\left( \mathbf{\nabla }\Phi \cdot \mathbf{\nabla }-\mathbf{\Delta }\right)
\left\langle \mathbf{\nabla }f(x_{o}),Mt_{1}\right\rangle \geq 0.
\end{equation*}%
This, then implies 
\begin{eqnarray*}
\left\langle \mathbf{\nabla }g(x_{o}),Mt_{1}\right\rangle &\geq
&\left\langle \mathbf{\nabla }f(x_{o}),\mathbf{Hess}\Phi
(x_{o})Mt_{1}\right\rangle \\
&=&\left\langle M\mathbf{\nabla }f(x_{o}),M^{-1}\mathbf{Hess}\Phi
(x_{o})Mt_{1}\right\rangle \\
&=&\left\langle t_{1},M^{-1}\mathbf{Hess}\Phi (x_{o})Mt_{1}\right\rangle \\
&\geq &\delta _{o}\left| \mathbf{\nabla }f(x_{o})\right| _{2,\rho }^{2}.
\end{eqnarray*}%
Thus%
\begin{eqnarray*}
\left| \mathbf{\nabla }f(x_{o})\right| _{2,\rho }^{2} &\leq &\frac{1}{\delta
_{o}}\left\langle M\mathbf{\nabla }g(x_{o}),t_{1}\right\rangle \\
&=&\frac{1}{\delta _{o}}\left\| M\mathbf{\nabla }g(x_{o})\right\| \left| 
\mathbf{\nabla }f(x_{o})\right| _{2,\rho }.
\end{eqnarray*}%
We have almost proved the following proposition

\begin{proposition}
\textit{Let }$g$\textit{\ be\ a smooth function satisfying}%
\begin{equation}
\left| \partial ^{\alpha }\mathbf{\nabla }g\right| \leq C_{\alpha
}\;\;\;\;\;\;\;\forall \alpha \in \mathbb{N}^{\left| \Gamma \right| }
\end{equation}%
and $\Phi $\textit{\ is as above. If }$f$ is \textit{the unique C}$^{\infty
}-$\textit{solution of the equation }%
\begin{equation*}
\left\{ 
\begin{tabular}{l}
$-\mathbf{\Delta }f+\mathbf{\nabla }\Phi \cdot \mathbf{\nabla }%
f=g-\left\langle g\right\rangle $ \\ 
$\left\langle f\right\rangle _{L^{2}(\mu )}=0,$%
\end{tabular}%
\right.
\end{equation*}%
\textit{\ then}%
\begin{equation*}
\sum_{i\in \Lambda }f_{x_{i}}^{2}(x)e^{2\kappa d(i,S_{g})}\leq C\;\;\ \
\forall x\in \mathbb{R}^{\Lambda }
\end{equation*}%
$C$ and $\kappa $\textit{\ are positive constants that could possibly depend
on the size of the support of }$g$\textit{\ but do not depend on }$\Lambda $%
\textit{\ and }$f.$
\end{proposition}

\begin{proof}
If%
\begin{equation}
\left| \mathbf{\nabla }f(x_{o})\right| _{2,\rho }=0
\end{equation}%
there is nothing to prove otherwise we have 
\begin{eqnarray*}
\left( \sum_{i\in \Lambda }f_{x_{i}}^{2}(x_{o})\rho ^{2}(i)\right) ^{1/2}
&\leq &\frac{1}{\delta _{o}}\left( \sum_{i\in \Lambda
}g_{x_{i}}^{2}(x_{o})\rho ^{2}(i)\right) ^{1/2} \\
&=&\frac{1}{\delta _{o}}\left( \sum_{i\in
S_{g}}g_{x_{i}}^{2}(x_{o})e^{2\kappa d(i,S_{g})}\right) ^{1/2} \\
&\leq &\frac{1}{\delta _{o}}\left( \sum_{i\in
S_{g}}g_{x_{i}}^{2}(x_{o})\right) ^{1/2}
\end{eqnarray*}%
and the result follows.
\end{proof}

\begin{corollary}
\textit{Let }$g$\textit{\ and }$h$\textit{\ be smooth functions on }$\mathbb{%
R}^{\Gamma },$\textit{and }$\mathbb{R}^{\Gamma ^{\prime }}$ \textit{where }$%
\Gamma $\textit{\ and }$\Gamma ^{\prime }\varsubsetneqq \Lambda $\textit{\
with }$\Gamma \cap \Gamma ^{\prime }=\varnothing $ \textit{denote
respectively the support of }$g$\textit{\ and }$h$\textit{\ and assume that }%
$g$ \textit{and }$h$\textit{\ satisfy }$\left( 4.22\right) .$\textit{\ Then
under the assumptions of proposition 2, we have }%
\begin{equation}
\left| \mathbf{cov}(g,h)\right| \leq Ce^{-\kappa d(S_{h},S_{g})}
\end{equation}%
\textit{where }$C$ and $\kappa $\textit{\ are positive constants that do not
depend on }$\Lambda ,$\textit{\ but possibly dependent on the size of the
supports of }$g$\textit{\ and }$h.$
\end{corollary}

\begin{proof}
Using the formula for the representation of the covariance, we have%
\begin{eqnarray*}
\left| \mathbf{cov}(g,h)\right| &=&\left| \left\langle A_{\Phi }^{1^{-1}}%
\mathbf{\nabla }g\mathbf{\cdot \nabla }h\right\rangle \right| \\
&=&\left| \left\langle \mathbf{\nabla }f\mathbf{\cdot \nabla }h\right\rangle
\right| \\
&\leq &\int \sum\limits_{i\in \Lambda }\left| f_{x_{i}}(x)e^{\kappa
d(i,S_{g})}e^{-\kappa d(i,S_{g})}h_{x_{i}}d\mu (x)\right| \\
&\leq &\int \left( \sum\limits_{i\in \Lambda }f_{x_{i}}^{2}(x)e^{2\kappa
d(i,S_{g})}\right) ^{1/2}\left( \sum\limits_{i\in
S_{h}}h_{x_{i}}^{2}(x)e^{-2\kappa d(i,S_{g})}\right) ^{1/2}d\mu (x) \\
&\leq &\left[ \int \sum\limits_{i\in \Lambda }f_{x_{i}}^{2}(x)e^{2\kappa
d(i,S_{g})}d\mu (x)\right] ^{1/2}\left[ \int \sum\limits_{i\in
S_{h}}h_{x_{i}}^{2}(x)e^{-2\kappa d(i,S_{g})}d\mu (x)\right] ^{1/2} \\
&\leq &C\left( \sum_{i\in S_{g}}g_{x_{i}}^{2}(x_{o})\right) ^{1/2}\left[
\int \sum\limits_{i\in S_{h}}h_{x_{i}}^{2}(x)d\mu (x)\right]
^{1/2}e^{-\kappa d(S_{h},S_{g})}.
\end{eqnarray*}
\end{proof}

\begin{remark}
This is the higher dimensional version of theorem 1.4 in [8]. Notice that
our proof does not require the assumptions $\left( 1.17\right) $ and $\left(
1.19\right) $ namely 
\begin{equation*}
\left\| \mathbf{Hess}\Phi (x)\right\| _{\mathcal{L}(l_{\rho }^{\infty
})}\leq C
\end{equation*}%
and 
\begin{equation*}
\left\| \mathbf{Hess}\Psi (x)\right\| _{\mathcal{L}(l_{\rho }^{\infty
})}\leq \rho <1
\end{equation*}%
for all $\rho $ as above. However, we required that $\Phi $ satisfies 
\begin{equation*}
M^{-1}\mathbf{Hess}\Phi (x)M\geq \delta _{o}
\end{equation*}%
for some $\delta _{o}\in (0,1)$ and $M$ as above.\newline
Notice also that the proof does not require any approximation of mean-field
type.
\end{remark}

\subsection{Relaxing the Compact Support Assumptions.}

We propose now to relax the assumptions of compact support made previously
on $\Psi $ and $g.$ As before, let $M$ be the diagonal matrix%
\begin{equation*}
M=\left( \delta _{ij}\rho (i)\right)
\end{equation*}%
where $\rho $ is given by \ 
\begin{equation}
\rho \left( i\right) =e^{\kappa d(i,S_{g})}
\end{equation}%
and 
\begin{equation}
M^{-1}\mathbf{Hess}\Phi (x)M\geq \delta _{o}
\end{equation}%
for every $M$ as above. Next, we propose to generalize the results in
propositions 22 without the assumptions of compact support on $\Psi $ and $g$
by means of a family of cutoff functions. Let us introduce as in [8] a
family cutoff functions 
\begin{equation}
\chi =\chi _{\varepsilon }
\end{equation}%
$(\varepsilon \in \lbrack 0,1])$ in $\mathcal{C}_{o}^{\infty }(\mathbb{R})$
with value in $[0,1]$ such that%
\begin{equation*}
\left\{ 
\begin{array}{c}
\chi =1\text{ \ \ \ \ \ \ \ \ \ \ \ \ \ \ \ \ for }\left| t\right| \leq
\varepsilon ^{-1}\text{ } \\ 
\left| \chi ^{(k)}(t)\right| \leq C_{k}\dfrac{\varepsilon }{\left| t\right|
^{k}}\text{ \ \ \ \ \ \ \ \ \ \ \ \ \ \ \ \ for }k\in \mathbb{N}.\mathbb{\ }%
\text{\ \ \ \ \ \ \ \ \ \ \ \ \ \ \ \ \ \ \ \ \ \ \ }%
\end{array}%
\right.
\end{equation*}%
We could take for instance%
\begin{equation*}
\chi _{\varepsilon }(t)=f(\varepsilon \ln \left| t\right| )
\end{equation*}%
for a suitable $f$. We then introduce%
\begin{equation}
\Psi _{\varepsilon }(x)=\chi _{\varepsilon }(\left| x\right| )\Psi ,\ \ \ \
\ \ \ \ \ x\in \mathbb{R}^{\Lambda }
\end{equation}%
and 
\begin{equation}
g_{\varepsilon }(x)=\chi _{\varepsilon }(\left| x\right| )g\text{ \ \ \ \ \
\ \ \ \ \ \ \ }x\in \mathbb{R}^{\Gamma }\text{.}
\end{equation}%
Recall that 
\begin{equation}
-\mathbf{\Delta }f+\mathbf{\nabla }\Phi \cdot \mathbf{\nabla }%
f=g-<g>_{t,\Lambda .}
\end{equation}%
which implies 
\begin{equation}
\left( -\mathbf{\Delta }+\mathbf{\nabla }\Phi \cdot \mathbf{\nabla }\right)
\otimes \mathbf{v}+\mathbf{Hess}\Phi \mathbf{v}=\mathbf{\nabla }g
\end{equation}%
where 
\begin{equation*}
\mathbf{v=\nabla }f.
\end{equation*}%
Under the transformations%
\begin{equation*}
\mathbf{v}=e^{-\Phi /2}\mathbf{u}\;\;\ \;\text{and \ \ \ \ }\mathbf{q}%
=e^{-\Phi /2}\mathbf{\nabla }g
\end{equation*}%
we have%
\begin{equation}
\left( \mathbf{-\Delta +}\frac{\left| \mathbf{\nabla }\Phi \right| ^{2}}{4}-%
\frac{\mathbf{\Delta }\Phi }{2}\right) \otimes \mathbf{Iu}+\mathbf{Hess}\Phi 
\mathbf{u}=\mathbf{q}\;\;\;\text{in }\mathbb{R}^{\Lambda }.
\end{equation}%
We first verify that the assumptions on $\Psi $ and $g$ are satisfied by $%
\Psi _{\varepsilon }(x)$ and $g_{\varepsilon }(x).$ Namely 
\begin{equation}
\left| \partial ^{\alpha }\mathbf{\nabla }\Psi \right| \leq C_{\alpha },\;\
\ \ \ \ \forall \alpha \in \mathbb{N}^{\left| \Lambda \right| },
\end{equation}%
\begin{equation}
\left| \partial ^{\alpha }\mathbf{\nabla }g\right| \leq C_{\alpha },\;\ \ \
\ \ \forall \alpha \in \mathbb{N}^{\left| \Lambda \right| },
\end{equation}%
and%
\begin{equation}
M^{-1}\mathbf{Hess}\Phi M\geq \delta >0,\;\;\;\text{ }0<\delta <1
\end{equation}%
$M$\ shall still denote the diagonal matrix%
\begin{equation*}
M=\left( \delta _{ij}\rho (i)\right) _{i,j\in \Lambda }
\end{equation*}%
where $\rho $ is a weight function on $\mathbb{R}^{\Lambda }$ satisfying \ 
\begin{equation}
e^{-\lambda }\leq \frac{\rho \left( i\right) }{\rho (j)}\leq e^{\lambda }%
\text{, \ \ if }i\sim j\text{\ \ for some }\lambda >0.
\end{equation}%
Using%
\begin{equation*}
\mathbf{Hess}\Psi \geq \delta -1,
\end{equation*}%
we obtain immediately 
\begin{equation}
M^{-1}\mathbf{Hess}\Psi _{\varepsilon }(x)M\geq \left( \delta -1\right) \chi
_{\varepsilon }(\left| x\right| )-C\varepsilon
\end{equation}%
for all $\varepsilon $ and some constant $C.$ Indeed,\ we know that 
\begin{equation*}
M^{-1}\mathbf{Hess}\Psi (x)M\geq \left( \delta -1\right) .
\end{equation*}%
For simplicity we shall write 
\begin{equation*}
\chi _{\varepsilon }=\chi \text{ \ and }r=\left| x\right|
\end{equation*}%
\begin{equation*}
\Psi _{\varepsilon }(x)=\chi (r)\Psi (x)
\end{equation*}%
\begin{eqnarray*}
\frac{\rho (j)}{\rho (i)}\Psi _{\varepsilon _{x_{i}x_{j}}} &=&\frac{1}{r}%
\frac{\rho (j)}{\rho (i)}\left( \delta _{ij}-\frac{x_{i}x_{j}}{r^{2}}\right)
\chi ^{\prime }(r)\Psi +\frac{\rho (j)}{\rho (i)}\frac{x_{i}x_{j}}{r^{2}}%
\chi ^{\prime \prime }(r)\Psi \\
&&+\frac{\rho (j)}{\rho (i)}\frac{x_{j}}{r}\chi ^{\prime }(r)\Psi _{x_{j}}+%
\frac{\rho (j)}{\rho (i)}\chi (r)\Psi _{x_{i}x_{j}}
\end{eqnarray*}%
Let $a\in \mathbb{R}^{\Lambda },$ 
\begin{eqnarray*}
&&\left\langle M^{-1}\mathbf{Hess}\Psi _{\varepsilon }(x)Ma,a\right\rangle \\
&=&\left( \frac{1}{r}\sum_{i}a_{i}^{2}-\frac{1}{r^{3}}\sum_{i,j}\frac{\rho
(j)}{\rho (i)}a_{i}a_{j}x_{i}x_{j}\right) \chi ^{\prime }(r)\Psi \\
&&+\frac{1}{r^{2}}\chi ^{\prime \prime }(r)\Psi \sum_{i,j}\frac{\rho (j)}{%
\rho (i)}a_{i}a_{j}x_{i}x_{j}+\frac{1}{r}\chi ^{\prime }(r)\sum_{i,j}\frac{%
\rho (j)}{\rho (i)}a_{i}a_{j}x_{j}\Psi _{x_{j}} \\
&&+\chi (r)\sum_{i,j}\frac{\rho (j)}{\rho (i)}a_{i}a_{j}\Psi _{x_{i}x_{j}} \\
&\geq &-2\frac{a^{2}}{r}\left| \chi ^{\prime }(r)\Psi (x)\right|
-a^{2}\left| \chi ^{\prime \prime }(r)\Psi (x)\right| -C\left| \chi ^{\prime
}(r)\right| a^{2}+\left( \delta -1\right) \chi (r)a^{2} \\
&\geq &\left[ \left( \delta -1\right) \chi (r)-\varepsilon C\right] a^{2}.
\end{eqnarray*}%
We conclude that \newline
\begin{equation*}
M^{-1}\mathbf{Hess}\Psi _{\varepsilon }(x)M\geq \left( \delta -1\right) \chi
(r)-\varepsilon C
\end{equation*}%
for all $\varepsilon >0.$\newline
It follows that 
\begin{equation}
M^{-1}\mathbf{Hess}\Phi _{\varepsilon }(x)M\geq \delta -C\varepsilon .
\end{equation}%
Now with $\delta $ replaced by $\delta ^{\prime }=\delta -C\varepsilon ,$ we
see that 
\begin{equation}
M^{-1}\mathbf{Hess}\Phi _{\varepsilon }(x)M\geq \delta ^{\prime
},\;\;\;\;\;\;\;\;0<\delta ^{\prime }<1
\end{equation}%
for $\varepsilon $ small enough. (Notice that $\varepsilon $ is possibly $%
\Lambda -$depend)It remains to check the assumptions on $g_{\varepsilon }$
and $\Psi _{\varepsilon }.$ To see that%
\begin{equation}
\left| \partial ^{\alpha }\mathbf{\nabla }g_{\varepsilon }\right| \leq C+%
\mathcal{O}_{\alpha ,\Lambda }(\varepsilon ),\;\ \ \ \ \ \forall \alpha \in 
\mathbb{N}^{\left| \Gamma \right| },
\end{equation}%
we have 
\begin{equation*}
g_{\varepsilon }(x)=\chi _{\varepsilon }(r)g(x),\text{ \ \ \ \ }x\in \mathbb{%
R}^{\Gamma }.
\end{equation*}%
Again let $\left| \alpha \right| \geq 1,$ using Leibniz's formula, we have 
\begin{eqnarray*}
\left| \partial ^{\alpha }g_{\varepsilon }\right| &\leq &\sum_{\beta \leq
\alpha }\dbinom{\alpha }{\beta }\partial ^{\beta }\chi _{\varepsilon
}(r)\partial ^{\alpha -\beta }g \\
&=&\left| \partial ^{\alpha }g\right| +\left| g\partial ^{\alpha }\chi
_{\varepsilon }(r)\right| +\sum_{\substack{ \beta <\alpha  \\ \beta \neq 0}}%
\dbinom{\alpha }{\beta }\left| \partial ^{\beta }\chi _{\varepsilon
}(r)\partial ^{\alpha -\beta }g\right| .
\end{eqnarray*}%
With the assumption $g(0)=0$, we write 
\begin{eqnarray}
\left| g(x)\right| &\leq &\int_{0}^{1}\sum_{j\in \Lambda }\left|
x_{j}g_{x_{j}}(sx)\right| ds \\
&\leq &\int_{0}^{1}\left( \sum_{j\in \Lambda }x_{j}^{2}\right) ^{1/2}\left(
\sum_{j\in \Lambda }g_{x_{j}}^{2}(sx)\right) ^{1/2}ds \\
&\leq &C_{g}r
\end{eqnarray}%
again using the fact that 
\begin{equation*}
r\partial ^{\alpha }\chi _{\varepsilon }(r)=\mathcal{O}_{\alpha
}(\varepsilon ),
\end{equation*}%
we get 
\begin{equation}
\left| g\partial ^{\alpha }\chi _{\varepsilon }(r)\right| =\mathcal{O}%
_{\alpha ,\Lambda }(\varepsilon ).
\end{equation}%
observe also that 
\begin{equation}
\sum_{\substack{ \beta <\alpha  \\ \beta \neq 0}}\dbinom{\alpha }{\beta }%
\left| \partial ^{\beta }\chi _{\varepsilon }(r)\partial ^{\alpha -\beta
}g\right| =\mathcal{O}_{\alpha }(\varepsilon )
\end{equation}%
it then immediately follows from the assumption on $g$ that 
\begin{equation}
\left| \partial ^{\alpha }\mathbf{\nabla }g_{\varepsilon }\right| \leq
C_{\alpha }+\mathcal{O}_{\alpha ,\Lambda }(\varepsilon ),\;\ \ \ \ \ \forall
\alpha \in \mathbb{N}^{\left| \Gamma \right| }.
\end{equation}%
Similarly, one can prove that 
\begin{equation}
\left| \partial ^{\alpha }\mathbf{\nabla }\Psi _{\varepsilon }\right| \leq
C_{\alpha }+\mathcal{O}_{\alpha ,\Lambda }(\varepsilon ),\;\ \ \ \ \ \forall
\alpha \in \mathbb{N}^{\left| \Lambda \right| },
\end{equation}%
Thus $\Psi _{\varepsilon }$ and $g_{\varepsilon }$ are compactly supported
and satisfy all the conditions that were previously required on $\Psi $ and $%
g$. If $\mathbf{u}_{\varepsilon }$ denotes the family of solutions
corresponding to the family of data $\Phi _{\varepsilon }$ and $%
g_{\varepsilon }$, one can see that $\mathbf{u}_{\varepsilon }$ converges to 
$\mathbf{u}$ in $C^{\infty }.$ The proof which based on regularity estimates
is given in detail in [8], Consequently, the family of solution $\mathbf{v}%
_{\varepsilon }=e^{\Phi _{\varepsilon }}\mathbf{u}_{\varepsilon }$ converges
to $\mathbf{v}$ in $C^{\infty }$

\begin{proposition}
\textit{If }$g(0)=0,$ \textit{then Proposition 22 holds without the
assumptions of compact support on }$\Psi $ \textit{and }$g.$
\end{proposition}

\begin{proof}
Using proposition 2 we have 
\begin{equation*}
\left( \sum_{i\in \Lambda }f_{\varepsilon _{x_{i}}}^{2}(x)e^{2\kappa
d(i,S_{g})}\right) ^{1/2}\leq C\left\vert S_{g}\right\vert ^{1/2}+\mathcal{O}%
_{\Lambda }(\varepsilon )\;\;\ \ \forall x\in \mathbb{R}^{\Lambda }.
\end{equation*}%
The result follows by taking the limit as $\varepsilon \rightarrow 0$
\end{proof}

\begin{corollary}
If $g=x_{i}$ and $h=x_{j}$ we get 
\begin{equation*}
\left| cor(i,j)\right| \leq Ce^{-\kappa d(i,j)}
\end{equation*}%
Which shows that we are away from a critical point.
\end{corollary}

\section{The d-dimensional Kac Model}

An example of a non-quadratic model satisfying the assumptions above is
given by 
\begin{equation*}
\Phi _{\Lambda }(x)=\frac{x^{2}}{2}-2\sum_{i\sim j}\ln \cosh \left[ \sqrt{%
\frac{\nu }{2}}\left( x_{i}+x_{j}\right) \right] .
\end{equation*}%
The summation is over all nearest neighbor sites.%
\begin{equation*}
\Psi (x)=-2\sum_{i,j\in \Lambda ,i\sim j}\ln \cosh \left[ \sqrt{\frac{\nu }{2%
}}\left( x_{i}+x_{j}\right) \right]
\end{equation*}%
with $\nu >0$ small enough.%
\begin{equation*}
\Psi _{x_{i}}=-2\sum_{j:,j\sim i}\frac{\sqrt{\frac{\nu }{2}}\sinh \left[ 
\sqrt{\frac{\nu }{2}}\left( x_{i}+x_{j}\right) \right] }{\cosh \left[ \sqrt{%
\frac{\nu }{2}}\left( x_{i}+x_{j}\right) \right] }
\end{equation*}%
\begin{equation*}
\Psi _{x_{i}x_{k}}=\left\{ 
\begin{tabular}{ll}
$-\nu \sum\limits_{j:,j\sim i}\dfrac{1}{\cosh ^{2}\left[ \sqrt{\frac{\nu }{2}%
}\left( x_{i}+x_{j}\right) \right] }$ & if $k=i$ \\ 
$-\dfrac{\nu }{\cosh ^{2}\left[ \sqrt{\frac{\nu }{2}}\left(
x_{i}+x_{k}\right) \right] }$ & if $k\sim i$ \\ 
$0$ & otherwise.%
\end{tabular}%
\right.
\end{equation*}%
It then follows that%
\begin{equation*}
\left| \Psi _{x_{i}}\right| \leq 4d\sqrt{\frac{\nu }{2}},
\end{equation*}%
\begin{equation*}
\left| \Psi _{x_{i}x_{i}}\right| \leq 2d\nu ,
\end{equation*}%
and 
\begin{equation*}
\left| \Psi _{x_{i}x_{k}}\right| \leq \nu \text{ \ \ \ \ if }k\sim i.
\end{equation*}%
Similarly, using the properties of $\cosh $ and $\sinh $ and the fact that $%
\sinh t\leq \cosh t$ for all $t$ one can see that all derivatives of order
greater than or equal to one are bounded. Now we propose to check that for $%
\nu $ small enough, the Kac Hamiltonian satisfies 
\begin{equation*}
M^{-1}\mathbf{Hess}\Phi (x)M\geq \delta _{o}
\end{equation*}%
for some $\delta _{o}\in (0,1)$ and $M$ as above.\newline
We need the following lemma.

\begin{lemma}[Schur's Lemma- The R and C bound]
For each rectangular array 
\begin{equation*}
\left( c_{ij}\right) _{\substack{ 1\leq i\leq m  \\ 1\leq j\leq n}}
\end{equation*}%
and each pair of sequence $\left( x_{i}\right) _{1\leq i\leq m}$ and $\left(
y_{j}\right) _{1\leq j\leq n}$ we have the bound 
\begin{equation*}
\left| \sum_{i=1}^{m}\sum_{j=1}^{n}c_{ij}x_{i}y_{j}\right| \leq \sqrt{RC}%
\left( \sum_{i=1}^{m}\left| x_{i}\right| ^{2}\right) ^{1/2}\left(
\sum_{j=1}^{n}\left| y_{j}\right| ^{2}\right) ^{1/2}
\end{equation*}%
where $R$ and $C$ are the row sum and column sum maxima defined by 
\begin{equation*}
R=\max\limits_{i}\sum_{j=1}^{n}\left| c_{ij}\right| \text{ \ \ \ \ and \ \ \
\ \ }C=\max\limits_{j}\sum_{i=1}^{m}\left| c_{ij}\right| .
\end{equation*}
\end{lemma}

This bound is known as Schur's Lemma, but, ironically, it may be the second
most famous result with this name. The Schur's decomposition lemma for $%
n\times n$ matrices is also known under this name. Nevertheless, this
inequality is surely the single most commonly used tool for estimating a
quadratic form. Going back to the example, we have for any $a=\left(
a_{i}\right) _{i\in \Lambda }\in \mathbb{R}^{\Lambda },$ 
\begin{eqnarray*}
&&\left\langle M^{-1}\mathbf{Hess}\Phi Ma,a\right\rangle  \\
&=&\sum_{i,j}\Phi _{x_{i}x_{j}}\frac{\rho (i)}{\rho (j)}a_{i}a_{j} \\
&=&\sum_{i}\Phi _{x_{i}x_{i}}a_{i}^{2}+\sum_{i\sim j}\Psi _{x_{i}x_{j}}\frac{%
\rho (i)}{\rho (j)}a_{i}a_{j} \\
&\geq &\left( 1-2d\nu \right) a^{2}+\sum_{i\sim j}\Psi _{x_{i}x_{j}}\frac{%
\rho (i)}{\rho (j)}a_{i}a_{j}.
\end{eqnarray*}%
Now using the Schur's lemma above, we have 
\begin{eqnarray*}
\left| \sum_{i\sim j}\Psi _{x_{i}x_{j}}\frac{\rho (i)}{\rho (j)}%
a_{i}a_{j}\right|  &\leq &\sum_{i,j}\left| \Psi _{x_{i}x_{j}}\frac{\rho (i)}{%
\rho (j)}a_{i}a_{j}\right|  \\
&\leq &\sqrt{RC}a^{2}
\end{eqnarray*}%
where 
\begin{equation*}
R=\max\limits_{i}\sum_{j}\left| \Psi _{x_{i}x_{j}}\frac{\rho (i)}{\rho (j)}%
\right| 
\end{equation*}%
and 
\begin{equation*}
C=\max\limits_{j}\sum_{i}\left| \Psi _{x_{i}x_{j}}\frac{\rho (i)}{\rho (j)}%
\right| .
\end{equation*}%
To estimate $R,$ observe that%
\begin{equation*}
\sum_{j}\left| \Psi _{x_{i}x_{j}}\frac{\rho (i)}{\rho (j)}\right| =\left|
\Psi _{x_{i}x_{i}}\right| +\sum_{j:j\sim i}\left| \Psi _{x_{i}x_{j}}\frac{%
\rho (i)}{\rho (j)}\right| .
\end{equation*}%
Now using the fact that 
\begin{equation*}
e^{-\kappa }\leq \frac{\rho (i)}{\rho (j)}\leq e^{\kappa }\text{ \ \ \ \ if }%
i\sim j
\end{equation*}%
we have 
\begin{equation*}
\sum_{j}\left| \Psi _{x_{i}x_{j}}\frac{\rho (i)}{\rho (j)}\right| \leq 2d\nu
+2d\nu e^{\kappa }.
\end{equation*}%
Hence 
\begin{equation*}
R\leq 2d\nu \left( 1+e^{\kappa }\right) .
\end{equation*}%
Similarly, we have 
\begin{equation*}
C\leq 2d\nu \left( 1+e^{\kappa }\right) .
\end{equation*}%
Thus, 
\begin{eqnarray*}
\left\langle M^{-1}\mathbf{Hess}\Phi Ma,a\right\rangle  &\geq &\left[ \left(
1-2d\nu \right) -2d\nu \left( 1+e^{\kappa }\right) \right] a^{2} \\
&=&1-2d\nu \left( 2-e^{\kappa }\right) .
\end{eqnarray*}%
The result follows by choosing $0<\kappa <\ln 2$ and $\nu <\dfrac{1}{%
2d\left( 2-e^{\kappa }\right) }.$

\textbf{Acknowledgements:} This work is part of my final thesis: Witten
Laplacian Methods for Critical phenomena.\textbf{\ }I would like to thank my
advisor Haru Pinson for all the fruitful discussions and the help he has
provided in the writing of these notes. I also would like to thank Prof. Tom
Kennedy, Prof. William Faris, and all members of the Mathematical Physics
group at the University of Arizona for their help and support.

\end{document}